\documentclass[jcp,amsmath,amssymb,showpacs,twocolumn]{revtex4-1}
\usepackage{graphicx,bm,amsmath,epsfig,ulem,color}
\pdfoutput=1

\def\noi{ \noindent }
\def\<{ \langle }
\def\>{ \rangle }

\begin{document}

\title{Hardening transition in a one-dimensional model for ferrogels}

 \author{ Mario Alberto Annunziata}
 \email{ annunziata@thphy.uni-duesseldorf.de }
 \author{ Andreas M.~Menzel }
 \email{ menzel@thphy.uni-duesseldorf.de }
 \author{ Hartmut L\"owen }
 \email{ hlowen@thphy.uni-duesseldorf.de }
 \affiliation{Institut f\"ur Theoretische Physik II, Heinrich-Heine-Universit\"at D\"usseldorf, Universit\"atsstra\ss e 1, D-40225 D\"usseldorf, Germany}

 \pacs{82.70.-y, 82.35.Np, 75.80.+q, 82.70.Dd}


\begin{abstract}
We introduce and investigate a coarse-grained model for quasi one-dimensional ferrogels. 
In our description the magnetic particles are represented by hard spheres with a magnetic dipole moment in their centers.
Harmonic springs connecting these spheres mimic the presence of a cross-linked polymer matrix. 
A special emphasis is put on the coupling of the dipolar orientations to the elastic deformations of the matrix, where a memory effect of the orientations is included. 
Although the particles are displaced along one spatial direction only, the system already shows rich behavior: 
as a function of the magnetic dipole moment, we find a phase transition between ``soft-elastic'' states with finite interparticle separation and finite compressive elastic modulus on the one hand, and ``hardened'' states with touching particles and therefore diverging compressive elastic modulus on the other hand. 
Corresponding phase diagrams are derived neglecting thermal fluctuations of the magnetic particles.
In addition, we consider a situation in which a spatially homogeneous magnetization is initially imprinted into the material. Depending on the strength of the magneto-mechanical coupling between the dipole orientations and the elastic deformations, the system then relaxes to a uniaxially ferromagnetic, an antiferromagnetic, or a spiral state of magnetization to minimize its energy. One purpose of our work is to provide a largely analytically solvable approach that can provide a benchmark to test future descriptions of higher complexity. From an applied point of view, our results could be exploited, for example, for the construction of novel damping devices of tunable shock absorbance.
\end{abstract}

\begin{widetext}
\thispagestyle{empty}
\noi \textit{Copyright 2013 American Institute of Physics. This article may be downloaded for personal use only. Any other use requires prior permission of the author and the American Institute of Physics.}

\vspace{.5cm}

\noi \textit{The following article appeared in The Journal Chemical Physics \textbf{138}, 204906 (2013) and may be found at\\ \url{http://jcp.aip.org/resource/1/jcpsa6/v138/i20/p204906_s1?isAuthorized=no}.}
\setcounter{page}{0}
\newpage
\end{widetext}

\maketitle


\section{Introduction}

Magnetic hybrid materials composed of ferro- or superparamagnetic particles in an elastic matrix have a 
variety of fascinating material properties \cite{filipcsei2007magnetic}.  
One important feature in view of possible applications arises from the fact that their elastic moduli  
can be tuned reversibly by applying an external magnetic field 
\cite{deng2006development,filipcsei2007magnetic,stepanov2007effect,chen2007investigation,bose2009magnetorheological}, 
which could be exploited for example to construct novel damping devices \cite{sun2008study} or vibration absorbers \cite{deng2006development}. 
The materials allow to combine specific properties of ferrofluids and magnetorheological fluids 
\cite{rosensweig1985ferrohydrodynamics,odenbach2003ferrofluids,odenbach2003magnetoviscous,huke2004magnetic,odenbach2004recent,fischer2005brownian,ilg2005structure,embs2006measuring,ilg2006structure,gollwitzer2007surface,vicente2011magnetorheological} 
with the convenience of crosslinked polymer systems: a confinement or container is not needed to maintain them. 

 Other properties studied for these systems are magnetostriction \cite{guan2008magnetostrictive} as well as deformations in nonuniform magnetic 
fields \cite{zrinyi1997direct,szabo1998shape}, swelling behavior under external magnetic fields \cite{filipcsei2010magnetodeformation}, shape 
memory effects \cite{nikitin2004magnetodeformational}, and their value for the design of soft actuators \cite{snyder2010design}. 
Furthermore, some of the properties induced by an external magnetic field can be permanently imprinted into the materials during the manufacturing process. 
For example, when during the crosslinking procedure an external magnetic field is applied that orients and arranges the magnetic particles, a significant 
degree of anisotropy can be locked in \cite{collin2003frozen,varga2003smart,gunther2012xray}. This anisotropy can show up in the optical, magnetic, 
and mechanical properties as well as in the swelling behavior. 

 By now, new preparation techniques and detailed characterization of the materials 
on the mesoscopic scale of the ferro-particles have lead to first results for a particle-resolved 
understanding of the structural and dynamical material properties. The degree of orientational coupling between the magnetic moments and the surrounding elastic matrix turns out 
to be a decisive parameter both from practical and theoretical points of view. 
This coupling is rather weak if the particles are small enough so that their magnetic moments can easily flip with respect to the particle 
axes, or if the particles themselves are only loosely incorporated into the polymer matrix and can easily rotate with respect to their environment \cite{zrinyi1996deformation,zrinyi1997ferrogel,krekhova2010thermoreversible}.
In contrary, this magneto-mechanical coupling is strong e.g.\ for bigger ferro-particles that are covalently bound to the polymer network 
\cite{bonini2008acrylamide,fuhrer2009crosslinking,frickel2011magneto,messing2011cobalt}. 

 Despite the importance of ferrogels in both applications and from a fundamental point of view as 
tunable soft matter, a complete theoretical description of their properties 
is missing. 
One difficulty arises from the long-ranged nature of the magnetic interactions, in particular when surface effects in samples of finite size are investigated \cite{weis2003simulation}. The long-ranged magnetic interactions are typically modeled using a dipolar form \cite{weis1993chain,hynninen2005phase,tavares1999dipolar}. For dipolar fluids the consequences of a finite system size and the presence of surfaces have been pointed out \cite{klapp2002spontaneous,klapp2005dipolar}. 
Another inherent problem is the simultaneous 
presence of many length scales ranging from the atomic or molecular scales of the polymer matrix over 
the mesoscopic scale of the magnetic particles to the macroscopic scales of a block of material. 
While there has been quite a number of previous theoretical works on the macroscopic scale
by using concepts from hydrodynamics \cite{jarkova2003hydrodynamics,bohlius2004macroscopic,bohlius2007solution} 
and elasticity theory \cite{stolbov2011modelling,zubarev2012theory}, there are only very few micro- and mesoscopic studies on ferrogels that explicitly incorporate the magnetic particles.

 One recent attempt has been performed by Camp and coworkers \cite{camp2011modeling,elfimova2012theory}, who also model the ferromagnetic particles as dipoles.
The dipolar pairwise interaction between neighboring 
particles is supplemented by a stabilizing hard core. The elastic matrix, however, 
 enters on the macroscopic level:
box shape fluctuations are penalized with an appropriate elastic energy contribution. 
Within this approach, affine deformations of the system are assumed. In other words, the macroscopic deformation of a block of material 
is affinely mapped to the displacement of each single particle. Consequently, the translational degrees of freedom of each particle are slaved 
to the superimposed macroscopic deformation. Ivaneyko et al.\ also used this simplifying assumption to study the behavior of paramagnetic or superparamagnetic particles
 that are rigidly positioned on a regular rectangular lattice \cite{ivaneyko2011magneto}. In their case the lattice as a whole can affinely deform on the 
cost of an elastic energy penalty that mimics the behavior of the crosslinked polymer matrix. When magnetic fields or mechanical deformations 
are externally imposed, the lattice affinely deforms to reduce the overall energy. Naturally rotations of the dipolar particle moments are not considered in Ref.~\cite{ivaneyko2011magneto} because the dipoles are 
induced by the static external magnetic field.

 At the next level of refinement, the translational degrees of freedom of each magnetic particle are included. In other words, a step beyond 
the assumption of affine deformations is made. Quite complementary to 
the modeling of Camp and coworkers \cite{camp2011modeling,elfimova2012theory}, the attempt by
Wojciechowski and coworkers \cite{dudek2007molecular} puts an effort in modeling the 
elastic matrix on the mesoscale level
but reduces the particles to effective beads ignoring their dipolar interaction.
The model of Wojciechowski and coworkers \cite{dudek2007molecular} has shown 
to reproduce---under suitable initial conditions---the negative Poisson ratio of 
dilational materials. Recently, Holm and coworkers \cite{weeber2012deformation} have introduced a microscopic dipole-spring
model for ferrogels which combines the essentials of both the elastic 
matrix and the dipolar nature of the particles. In Ref.\ \cite{weeber2012deformation}, a finite two-dimensional block of material 
was simulated at finite temperature in order to get a first insight into the macroscopic properties of ferrogels. 

 More refined dipole-spring approaches have been employed very recently to study the conformation of flexible magnetic filaments by S\'anchez, Cerd\`a, and coworkers \cite{sanchez2013filaments,cerda2013filaments}. In the first work \cite{sanchez2013filaments}, they model a single supramolecular magnetic filament as a bead-spring chain of freely rotating dipoles. The existence of different conformations of the chains in three spatial dimensions is demonstrated by varying the strength of the dipolar interaction. A similar model is used in Ref.~\cite{cerda2013filaments}, but there a coupling of the dipolar orientations to the orientation of the filament axis is additionally included. The phase diagram of the equilibrium conformations of a single flexible magnetic chain in a poor solvent is derived. It features compact, helicoidal, partially collapsed, simply closed, and extended open states as a function of the temperature and of the relative strength of the interparticle interactions.

 In the present paper, we consider a quasi-1d ferrogel model system of hard-sphere particles of equal magnetic dipolar moment. 
Its phase behavior is investigated in the presence of an external magnetic field with special focus on the rotational coupling between 
the dipolar moments and the elastic polymer matrix. For this purpose we include the dipole-dipole interaction between particles and an explicit dipolar-matrix interaction. The system is quasi-1d because the particles are displaced along one direction of space only, whereas the magnetic moments can reorient in all spatial directions. 
We propose a coarse-grained dipolar-spring approach, which is to a big extent analytically solvable. 
Neglecting thermal fluctuations, i.e.\ considering a state of zero temperature for the magnetic particles, this model already in 1d exhibits a phase transition between states of zero (``hardened'') and finite (``soft-elastic'') particle-separation.
The corresponding phase diagram includes a critical point. We also take into account the possibility of an orientational memory concerning the dipole orientations with respect to their environment. In this case the properties of the ``soft-elastic''--``hardened'' transition change only quantitatively. However, qualitatively different energetic ground state solutions can emerge as a consequence. 

 To illustrate this point, we consider the following situation. A sample is generated under the influence of a strong external magnetic field that homogeneously orients all dipolar moments into the same direction obliquely to the 1d-system axis. The orientational memory of this initial state is permanently imprinted into the system. When the external magnetic field is switched off, the system minimizes its energy under the influence of the orientational memory. 
The picture varies sensibly as a function of the magnitude of the magnetic dipole moment as well as of the strength of the orientational memory. We find that for small values of the magnetic moment or strong orientational memory, the new energetic ground state is given by a spiral arrangement of the dipolar moments around the system axis. For larger values of the magnetic moment and weaker orientational memory, instead, we can find one of the following two states minimizing the energy of the system: a uniaxial ferromagnetic configuration in which the dipolar moments are aligned along the system axis; or an alternating antiferromagnetic-like state in which the relative orientations of neighboring dipoles are given by half turns around the system axis.

 Our motivation to study a one-dimensional model is twofold:
first, as is known in general in statistical mechanics, any
analytical solution of a one-dimensional model may serve as a benchmark
to test general approximate theories and thereby provide constraints
to the theory. One important example is the classic
Tonks gas of hard rods in one dimension
  where the full density functional is analytically
soluble \cite{Percus}. This has been found to be a key element in 
constructing
approximative density functionals in higher dimensions via the
dimensional crossover constraint \cite{rosenfeld1997fundamental}.
The second motivation is to include effects of orientational memory of the
ferrogel which have not yet been studied before. Therefore
it is useful to explore them within a simple one-dimensional model first 
before generalizing them to higher dimensions.

\vspace{5pt}

 The paper is organized as follows. In Sec.~\ref{sec2} we present our model and give the basic definitions.
We discuss in Sec.~\ref{sec3} a situation without orientational memory but in the presence of a strong external magnetic field, while in Sec.~\ref{sec4} the phase diagram and the critical-point
positions are determined under the influence of an orientational memory. In the latter case, qualitatively different ground state solutions are found depending on the strength of the magneto-mechanical coupling. Finally, our conclusions are given in Sec.~\ref{sec5}.

\section{The model} \label{sec2}

In the following we consider a ferrogel system in a thin cylinder. The ferro-particles are distributed along the cylindrical axis and the remaining space 
in the cylinder is filled by the polymer such that the arrangement of the ferro-particles is one-dimensional. Either ferromagnetic particles or, with a small modification \cite{froltsov2003crystal}, superparamagnetic particles are described by our model. 

 In real materials, two different situations can be realized nowadays \cite{frickel2011magneto,messing2011cobalt}: 
either the particles are physically caged within the mesh pockets of the enclosing chemically crosslinked polymer network and can rotate within these mesh pockets; 
or the particles are permanently covalently bound into the polymer network so that they become themselves a part of the network upon crosslinking. The latter feature can be achieved through surface-functionalization of the particles \cite{frickel2011magneto,messing2011cobalt}. 
During the crosslinking process a strong magnetic field can be applied that aligns the ferromagnetic particles with their easy axis and magnetic moment along the field direction while the crosslinking of the polymer chains is taking place. As a consequence, even when the field 
is turned off, the particles ``remember'' their original orientation. Therefore, at least in the situation of covalently bound particles, an orientational memory of the dipolar orientations during the crosslinking process is generated \cite{collin2003frozen}.

\begin{figure}
\begin{center} 
\includegraphics[width=8.5cm]{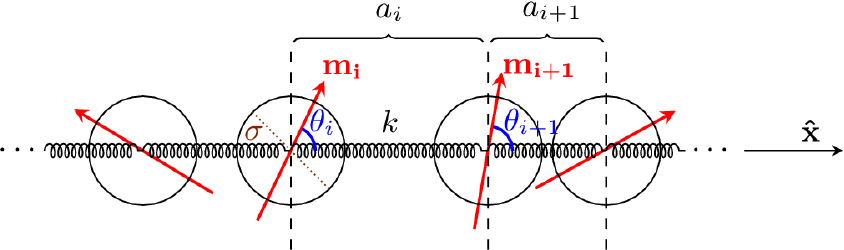}
\caption{(Color online) 
Schematic figure of the model: every two neighboring particles $i$ and $i + 1$ ($i = 1, \cdots, N - 1$) are connected by a spring of elastic constant $k$ 
attached to their centers, the center-to-center distance being $a_i$. Each particle is represented by a hard sphere of diameter $\sigma$, and each magnetic moment ${\bf m}_i$ ($i = 1, \cdots, N$) forms the angle $\theta_i$ with the chain axis $\mathbf{\hat{x}}$.
}
\label{fig0}
\end{center}
\end{figure}

 We model the quasi-1d ferrogel system as a 1d-chain made of $N$ hard spherical particles of diameter $\sigma$ and dipolar magnetic moments ${\bf m}_i$. Here, the index $i = 1, \cdots, N$ labels the particles. The orientation of each magnetic moment $\mathbf{m}_i$ is parameterized by the two angles $\theta_i$ and $\phi_i$. $\theta_i$ corresponds to the angle between $\mathbf{m}_i$ and the chain axis, whereas $\phi_i$ measures the azimuthal orientation around the chain axis. Neighboring particles are connected by springs having an initial length $L$ and an elastic constant $k$, as illustrated in Fig.~\ref{fig0}. 
In this first approach we will completely neglect the effect of thermal fluctuations on the magnetic particles. Our treatment in the following will therefore be a purely energetic one.

The elastic part of our energy is 
\begin{equation}\label{eq:h_el}
 E_{el} = \frac{k}{2} \sum_{\< i,j \>} \left( r_{ij} - L \right)^2.
\end{equation}
 where $r_{ij}=\|\mathbf{r}_{ij}\|$ with ${\bf r}_{ij} = {\bf r}_j - {\bf r}_i$. Here, the angular brackets $\langle i, j \rangle$ denote a sum that is evaluated only over nearest neighbors. The dipole-dipole interaction is given by
\begin{equation}\label{eq:h_dip_0}
 E_{dip} = \frac{\mu_0}{4 \pi} \sum_{i,j=1, i < j}^N \frac{({\bf m}_i \cdot {\bf m}_j)r^2_{ij} - 3({\bf m}_i \cdot {\bf r}_{ij})({\bf m}_j \cdot {\bf r}_{ij})}{r^5_{ij}}.
\end{equation}
In the case of superparamagnetic particles magnetized by an external magnetic field an additional factor of $1/2$ enters this formula \cite{froltsov2003crystal}. It can be considered by rescaling the constant $\mu_0$ that will only appear in this energetic contribution. 

 We introduce a rotational coupling term between the magnetic dipoles and the distance vectors via
\begin{equation}\label{eq:h_D}
 E_D = D \sum_{\< i,j \>} \left( \frac{{\bf m}_i \cdot {\bf r}_{ij}}{  m_i \: r_{ij} } - \frac{{\bf m}_i^{(0)} \cdot {\bf r}_{ij}^{(0)}}{ m_i^{(0)}  r_{ij}^{(0)} } \right)^2,
\end{equation}
where $D$ sets the strength of the coupling, ${\bf m}_i^{(0)}$ and ${\bf r}_{ij}^{(0)}$ are, respectively, the initial magnetic moment and the initial relative distance vectors after cross-linking, and $m_i=\|\mathbf{m}_i\|$, $m_i^{(0)}=\|\mathbf{m}_i^{(0)}\|$, as well as $r_{ij}^{(0)}=\|\mathbf{r}_{ij}^{(0)}\|$. In short, this contribution means that rotations of the magnetic dipole moments towards or away from the chain axis cost energy. Such a rotation has occurred, if the new dipole orientation ${\bf m}_i/m_i$ with respect to the direction ${\bf r}_{ij}/r_{ij}$ differs from the initial dipole orientation ${\bf m}_i^{(0)}/m_i^{(0)}$ with respect to the initial direction ${\bf r}_{ij}^{(0)}/r_{ij}^{(0)}$. Consequently, we can include via this term an orientational memory. It becomes important when we model the case of ferromagnetic particles that are covalently bound to the surrounding crosslinked polymer network.

\begin{figure}
\begin{center} 
\includegraphics[width=6.5cm]{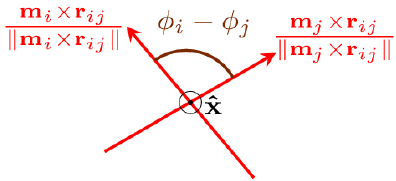}
\caption{(Color online) 
Schematic figure of the model for two neighboring particles $i$ and $j$ ($i = 1, \cdots, N - 1$, $j = i + 1$) as seen from along the chain axis $\mathbf{\hat{x}}$ ($\mathbf{\hat{x}}$ is oriented towards the reader). For the $i$-th particle the direction of the projection of the magnetic moment into the plane perpendicular to the chain axis is given by ${\bf m}_i \times {\bf r}_{ij}/\| {\bf m}_i \times {\bf r}_{ij} \|$, except for an additional rotation by $\pi/2$. The azimuthal orientation of each magnetic moment ${\bf m}_i$ ($i = 1, \cdots, N$) around the chain axis $\mathbf{\hat{x}}$ is measured by the angle $\phi_i$.}
\label{fig0a}
\end{center}
\end{figure}

 In the latter case, elastic torsional deformations of the polymer network must also be taken into account. These occur when different ferromagnetic particles together with their dipolar moments rotate around the chain axis by different angles. To quantify this situation, neighboring magnetic moments are projected into a plane perpendicular to the chain axis; then we focus on the angle between these projections of neighboring dipolar moments. If this angle has changed between the current configuration and the initial configuration, an energetic penalty arises. This whole procedure is achieved through the following energetic contribution: 
\begin{eqnarray}
 E_\tau &=& \tau \sum_{\< i,j \>} \Bigg(  \frac{{\bf m}_i \times {\bf r}_{ij}}{\| {\bf m}_i \times {\bf r}_{ij} \| } \cdot \frac{{\bf m}_j \times {\bf r}_{ij}}{\| {\bf m}_j \times {\bf r}_{ij} \|  }  \nonumber\\ &&\quad{}
-  \frac{{\bf m}_i^{(0)} \times {\bf r}_{ij}^{(0)}}{\| {\bf m}_i^{(0)} \times {\bf r}_{ij}^{(0)} \| } \cdot \frac{{\bf m}_j^{(0)} \times {\bf r}_{ij}^{(0)}}{\| {\bf m}_j^{(0)} \times {\bf r}_{ij}^{(0)} \|}  \Bigg)^2, \label{eq:h_tau}
\end{eqnarray}
where $\tau$ gives the strength of the coupling. The orientation of the projection of $\mathbf{m}_i$ into the plane perpendicular to the chain axis is given by the fraction term ${\bf m}_i \times {\bf r}_{ij}/\| {\bf m}_i \times {\bf r}_{ij} \|$ (apart from an additional rotation by $\pi/2$), and likewise for $\mathbf{m}_j$, see Fig.~\ref{fig0a}. The scalar product between the projections of $\mathbf{m}_i$ and $\mathbf{m}_j$ gives a measure for the difference in azimuthal angles $\phi_i - \phi_j$ between neighboring projected orientations. Then the difference between the current and the initial configuration is taken by subtracting the two scalar products. 

 Finally, the hard-sphere interaction between the magnetic particles reads
\begin{equation}
 E_{\rm HS} = \sum_{\< i,j \>} u_{\rm HS} (r_{ij}),
\end{equation}
 where $u_{\rm HS} (r_{ij})$ is the two-body hard-sphere potential
\begin{equation}
 u_{\rm HS} (r) = \left\{ 
 \begin{array}{ll}
  + \infty & {\rm if} \qquad r < \sigma \\
  0 & {\rm if} \qquad r \geq \sigma
 \end{array}
 \right. ,
\end{equation}
with $\sigma$ the diameter of the particles. This stabilizes the system, because it avoids a collapse when the dipole-dipole interaction is attractive. 

 In a situation without orientational memory, for example for superparamagnetic particles, or for ferromagnetic particles that are not covalently bound into the surrounding polymer matrix and can relatively easily rotate within their polymer mesh pockets, the energy is given by 
\begin{equation}\label{eq:ham_1}
 E_1 \equiv E_{el} + E_{dip} + E_{\rm HS}.
\end{equation}
In this case, we will assume that the orientation of the magnetic moments is fixed from outside. Such a situation can occur when the magnetic particles are exposed to a strong homogeneous external magnetic field and can easily rotate with respect to their environment, or when the particles are superparamagnetic \cite{froltsov2003crystal,froltsov2005anisotropic}.

 Conversely, when orientational memory is important, for example when ferromagnetic particles are covalently crosslinked into the surrounding polymer network, the energy reads
\begin{equation}\label{eq:ham_2}
 E_2 \equiv E_1 + E_D + E_\tau.
\end{equation}

 From now on, we assume that the magnitude of the magnetic moment is the same for all particles $m_i=m$, $i=1,...,N$. Thus ${\bf m}_i = m {\bf \hat{u}}_i$, $i=1,...,N$, with the unit vectors ${\bf \hat{u}}_i$ giving the orientations of the magnetic moments.

\section{No orientational memory} \label{sec3}

In this section we study the behavior of the system described by the energy $E_1$ of Eq.~(\ref{eq:ham_1}). Since there is no orientational memory in $E_1$, 
this corresponds to the case of superpara\-magnetic particles, the magnetic moments of which can rotate with respect to the particle axes, or to the case of ferromagnetic particles that can rotate with respect to the polymer matrix without energetic penalty on the 
considered time scales. We start by considering a strong external magnetic field. Then the orientation angle of the dipoles can be viewed as an externally fixed parameter 
$\theta_i = \theta_B$ ($i = 1, \cdots, N$), where $\theta_B$ is the angle that the magnetic field forms with the chain.

Looking for spatially homogeneous solutions only, we assume that all the interparticle distances $a_i$ are equal, i.e.~$a_i = a$ for $i = 1, \cdots, N$. For simplicity the thermodynamic limit of large $N$ is considered. As a consequence the energy per particle becomes
\begin{equation}\label{eq:h1}
 \frac{E_1}{N} = \frac{k}{2} (a - L)^2 + \frac{\mu_0 \zeta(3) m^2}{4 \pi a^3}  (1 - 3 \cos^2 \theta_B) + \frac{E_{\rm HS}}{N},
\end{equation}
 where $\zeta(3) = \sum\limits_{n = 1}^\infty {\frac{1}{n^3}} \approx 1.20206$ and $\zeta$ is the Riemann zeta-function. We assume that the distance $a$ can freely adjust itself to minimize this energy, or, in other words, that there is no external pressure applied. Minimizing Eq.~(\ref{eq:h1}) with respect to $a$, the interparticle distance $a^*$ defined by $d(E_1/N)/da|_{a*}=0$ is obtained from the implicit condition
\begin{equation}\label{eq:astar}
 {a^*}^4 \left( a^*-L \right) - \frac{3 \mu_0 \zeta(3)}{4 \pi k} m^2 \left(1 - 3 \cos^2 \theta_B \right) = 0.
\end{equation}

 The corresponding elastic modulus $G$,
\begin{equation}
 G(a^*) \equiv \frac{1}{N} \left( \frac{d^2 E_1}{da^2} \right)_{a^*},
\end{equation}
 becomes
\begin{equation}\label{eq:G}
 G(a^*) = k \left[ 1 + 4\left( 1 - \frac{L}{a^*} \right) \right].
\end{equation}

Both quantities, $a^*$ and $G(a^*)$, can be radically modified by varying the angle of the external magnetic field $\theta_B$ because at 
\begin{equation}\label{theta1}
 \theta_1 \approx 0.3 \pi \approx 54.7^{\circ}, \quad \theta_2 \approx 0.7 \pi \approx 125.3^{\circ}
\end{equation}
 the dipolar interaction vanishes. For $\theta_B \in [0,\theta_1) \cup (\theta_2,\pi]$ the dipolar interaction is attractive, while for $\theta_B \in (\theta_1,\theta_2)$ it is repulsive. The attractive case is particularly interesting, because for each $\theta_B$, a critical magnetic moment $m_c$ exists such that $G = 0$. Furthermore, for $m > m_c$ the elastic compressibility modulus $G$ diverges. We call this the ``hardened'' phase, in which the hard spheres touch each other so that the system cannot be compressed any further. From the condition $G = 0$ the critical values of the interparticle distance and of the magnetic moments are obtained as
\begin{eqnarray}
 &a^*_c& = \frac{4}{5} L, \label{eq:ac} \\
 &m_c (\theta_B)& = \frac{16}{25} \sqrt{\frac{4 \pi}{15 \zeta(3) (3 \cos^2 \theta_B - 1)}} m_0, \label{eq:mc}
\end{eqnarray}
where $m_0 \equiv \sqrt{k L^5 / \mu_0}$ provides a unit of measurement for the magnetic moment. 
It is interesting to note that, at variance with the critical value of the magnetic moment $m_c$, the critical value of the interparticle distance $a^*_c$ does not depend on the angle of the external magnetic field $\theta_B$, i.e.~it is a purely mechanical feature of the model. Furthermore, $a^*_c$ is solely determined by the initial particle distance $L$, whereas the hard sphere diameter $\sigma$ does not enter the expression. 
Since here we assume spatial homogeneity together with an identical orientation of all magnetic moments, the dipolar moment $m$ of the single particles is proportional to the overall magnetization of the sample.

\begin{figure}
\begin{center} 
\includegraphics[width=8.5cm]{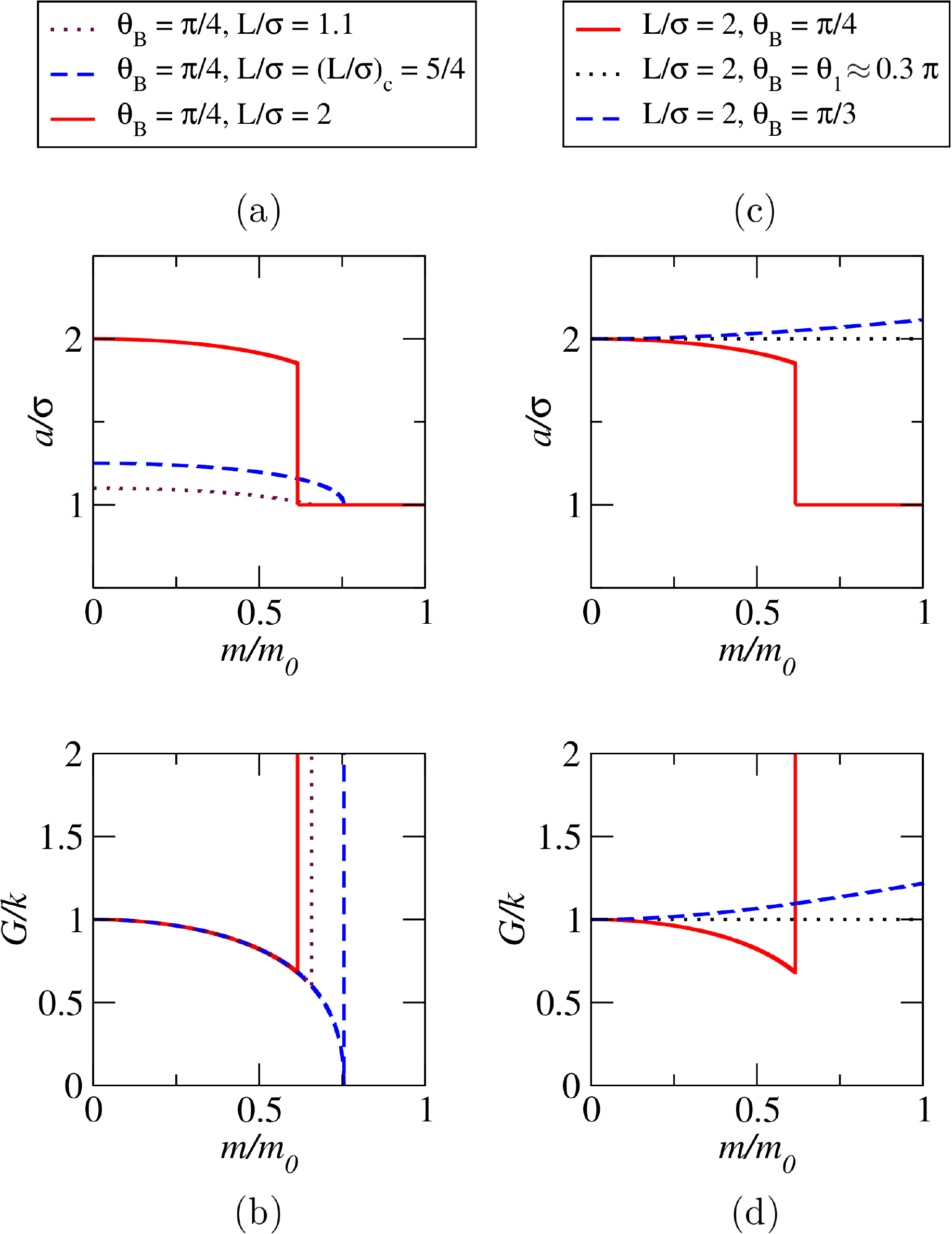}
\caption{(Color online) 
Interparticle distance $a/\sigma$ and compressive elastic modulus $G$ at equilibrium, as a function of the magnetic dipole moment $m/m_0$. In (a) and (b) the three different lines correspond to three different values of the rescaled initial particle separation $L/\sigma$, whereas the orientation of the external magnetic field with respect to the system axis is kept fixed at $\theta_B = \pi / 4$; here $L/\sigma = 1.1$ (dotted line), $L/\sigma = (L/\sigma)_c = 5/4$ (dashed line), and $L/\sigma = 2$ (solid line). In (c) and (d) the three different lines correspond to three different orientations $\theta_B$ of the external magnetic field with respect to the system axis, whereas the rescaled initial particle separation is kept fixed at $L/\sigma = 2$; here $\theta_B = \pi / 4$ (solid line), $\theta_B = \theta_1\approx 0.3\pi$ (dotted line), and $\theta_B = \pi / 3$ (dashed line).}
\label{fig1}
\end{center}
\end{figure}

 In the following we use the ratio $m/m_0$ as a control parameter. We illustrate the equilibrium behavior of $a(m/m_0)$ and $G(m/m_0)$ in Fig.~\ref{fig1}. On the left-hand side, we compare the behavior of $a(m/m_0)$ and $G(m/m_0)$ for different rescaled initial interparticle distances $L/\sigma$, but for the same orientation angle of the external magnetic field $\theta_B=\pi/4$. Comparing to the angle $\theta_1$ in Eq.~(\ref{theta1}), we find that the dipolar interaction is attractive in this case. Therefore the interparticle distance $a>\sigma$ always decreases when the magnetic moment $m/m_0$ is increased as depicted in panel (a), until the particles touch each other, $a=\sigma$. For a rescaled initial particle separation $L/\sigma<(L/\sigma)_c$ (dotted line) this transition to the touching state is continuous. For $L/\sigma>(L/\sigma)_c$ (solid line) the transition is discontinuous. The behavior at the critical value $(L/\sigma)_c=5/4$ (dashed line) is also shown. Of course, when the hard spheres touch each other, the compressive elastic modulus $G$ diverges, as given for all cases in panel (b).  On the right-hand side of Fig.~\ref{fig1} we compare the different behaviors of $a(m/m_0)$ and $G(m/m_0)$ for the same rescaled initial particle separation $L/\sigma$, but for different orientation angles of the external magnetic field $\theta_B$. In the first case, $\theta_B=\pi/4<\theta_1$ (solid line), the dipolar interaction is attractive. Again the same curves (solid lines) as in panels (a) and (b) are shown for a discontinuous transition to the ``hardened'' state of touching particles $a=\sigma$ and diverging compressive elastic modulus $G$. If the dipolar interaction is repulsive instead, as depicted here for $\theta_1<\theta_B=\pi/3<\theta_2$ (dashed line), the particles tend to separate from each other with increasing $m/m_0$ due to the growing repulsion. Consequently we find a growing $a$ in panel (c). Still, however, we observe that the modulus $G$ increases with the magnetic moment as depicted in panel (d). In the neutral case between dipolar attraction and repulsion, $\theta_B=\theta_1$ (dotted line), both the particle separation $a$ and the modulus $G$ remain constant when $m/m_0$ is changed.

\begin{figure} 
\begin{center} 
\includegraphics[width=8.5cm]{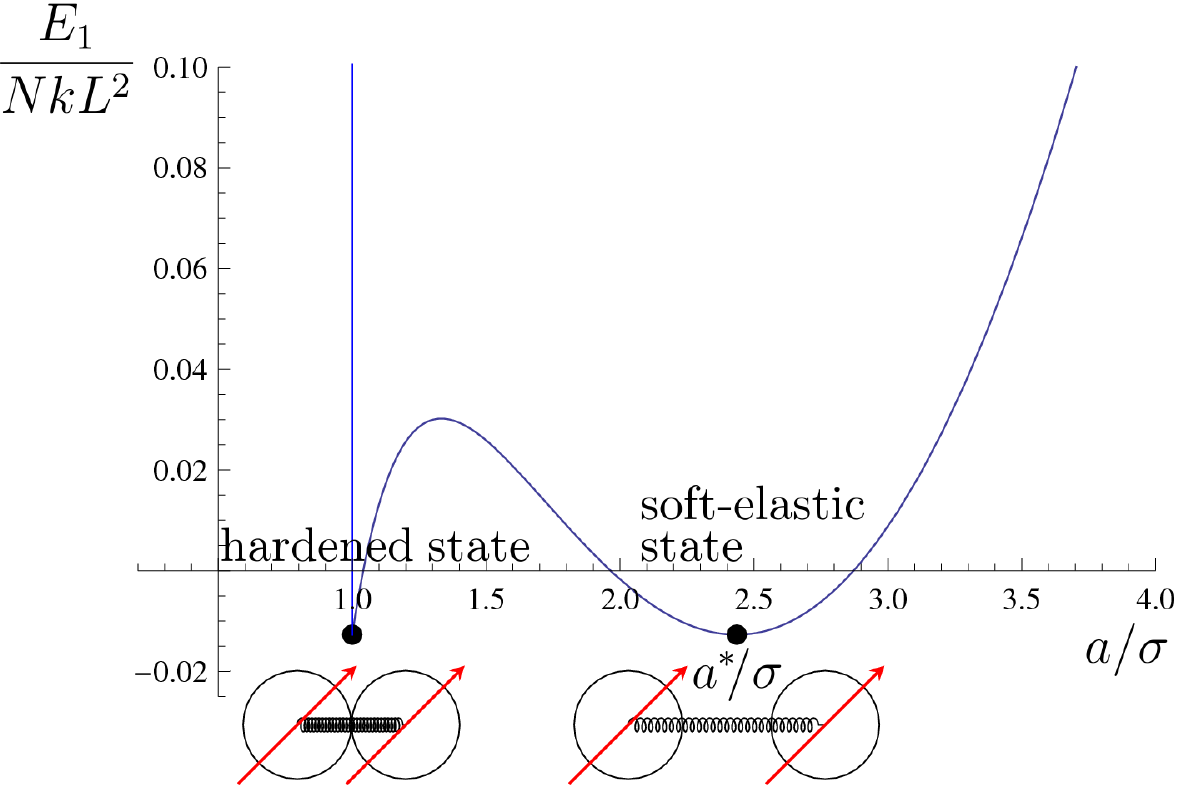}
\caption{(Color online) 
Energy per particle $E_1/N$ as a function of the rescaled interparticle distance $a/\sigma$, for $m/m_0 = 0.5$ and $\theta_B = \pi/4$. The black points are equal-energy minima of $E_1/N$: at the point $a/\sigma=1$ the hard spheres touch each other and the system is in a ``hardened'' state, while at the point $a/\sigma = a^*/\sigma \approx 2.45 $ the system is still ``soft-elastic''.}
\label{fig2}
\end{center}
\end{figure}

Fig.~\ref{fig2} shows an example of the energy per particle $E_1/N$ as a function of the rescaled interparticle distance $a/\sigma$. 
In this case the energy can exhibit two local minima. The first one is located at the point $a^*/\sigma$ defined by $d(E_1/N)/da|_{a*}=0$. Here the particles are elastically connected to each other and separated by a finite distance. When a compressive force is applied, the system can be deformed reversibly, so that we call this state ``soft-elastic''. The second minimum 
is located at the intersection point of the $E_1/N$ curve with the line $a / \sigma=1$. Now there is no gap left between the particles so that the hard spheres touch each other. Consequently the elastic compressibility modulus $G$ diverges because the system cannot be compressed any further. By varying $m/m_0$ we can tune the energy corresponding to these two minima to be equal. This special situation realizes two equilibrium states with the same energy, i.e.~a phase coexistence between the ``hardened'' state $a / \sigma=1$ and the ``soft-elastic'' state $a /\sigma= a^*/\sigma$. Since these two phases are characterized by a different value of the interparticle distance, we consider the equilibrium particle distance $a^* / \sigma - 1$ as an order parameter. The control parameters are the rescaled initial particle separation $L/\sigma$ and the magnetic moment $m/m_0$.

\begin{figure} 
\begin{center} 
\includegraphics[width=8.5cm]{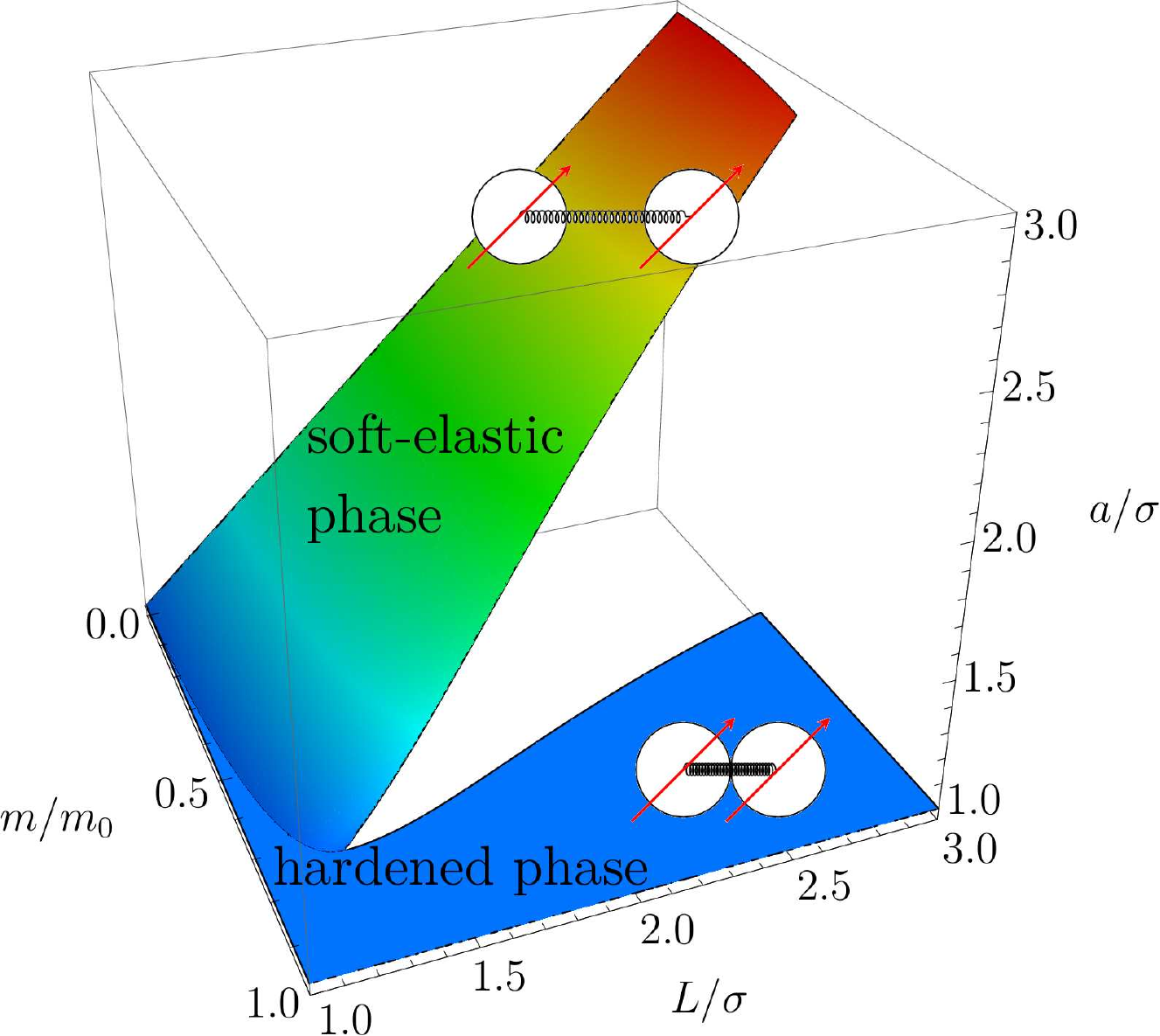}
\caption{(Color online) 
Phase diagram for a quasi-1d ferrogel system without orientational memory in the presence of a strong external magnetic field that is tilted with respect to the system axis by the angle $\theta_B = \pi/4$. The bottom plane $a / \sigma=1$ corresponds to the ``hardened'' phase, while the upper tilted surface corresponds to the ``soft-elastic'' phase. For $1 < L/\sigma < 5/4$ the transition is continuous, while it is discontinuous for $L/\sigma > 5/4$.}
\label{fig3}
\end{center}
\end{figure}

As a function of the magnetic moment $m/m_0$, we find a phase transition between these two states. This transition is discontinuous for $L/\sigma > 5/4$ and becomes continuous for $1 < L/\sigma < 5/4$, with a second-order critical point at $(L / \sigma)_c = 5/4$, as illustrated in Fig.~\ref{fig1} (a) and (b). In Fig.~\ref{fig3} we show the phase diagram of the system, here for $\theta_B = \pi/4$. Within the phase diagram the location of the critical point is given by the coordinates 
\begin{eqnarray}
(L/\sigma)_c & = & \frac{5}{4},   \label{coordinates_cpL}
\end{eqnarray}
\begin{eqnarray}
(m/m_0)_c & = &  \frac{16}{25}\sqrt{\frac{4\pi}{15\zeta(3)(3\cos^2\theta_B-1)}}, 
  \label{coordinates_cpm}\\
(a/\sigma)_c & = & 1.   \label{coordinates_cpa}
\end{eqnarray}
While the plane $a / \sigma=1$ corresponds to the ``hardened'' phase, the upper tilted surface corresponds to the ``soft-elastic'' phase. We emphasized in the figure the continuous nature of the transition for $1 < L/\sigma < 5/4$ and its discontinuous nature for $L/\sigma > 5/4$.

The phase transition at the critical point given by Eqs.~(\ref{eq:ac}) and (\ref{eq:mc}) is of second order. From Eq.~(\ref{eq:astar}) we obtain the magnetostrictive susceptibility as
\begin{equation}\label{eq:der}
 \chi \equiv \frac{d a^*}{dm} = \frac{3 \mu_0 \zeta(3) m (1 - 3 \cos^2 \theta_B)}{10 \pi k {a^*}^3 (a^* - a^*_c)},
\end{equation}
 which diverges for $a^* \to a^*_c$ [see Eq.~(\ref{eq:ac})]. Furthermore, we can extract the critical  exponent of $\chi$ when the magnetic moment and the rescaled interparticle distance approach their critical values $(m/m_0)_c$ and $(a/\sigma)_c$, respectively,  as given by Eqs.~(\ref{coordinates_cpL})--(\ref{coordinates_cpa}). This limit is taken at fixed $L/\sigma = (L/\sigma)_c=5/4$, corresponding to the dashed curve of Fig.~\ref{fig1} (a). The exponent is defined by 
\begin{equation}
 \chi \sim \frac{1}{a^* - a^*_c} \sim  |m/m_0 - (m/m_0)_c|^{-\beta}. 
\end{equation}
 We find $\beta = 1/2$ as it is typical for any mean-field approach. This result can be obtained analytically by integrating Eq.~(\ref{eq:der}) and neglecting the term ${a^*}^3 \to {a^*_c}^3$ for $m/m_0 \to (m/m_0)_c$.

\section{Orientational memory} \label{sec4}

In this section we study the behavior of the system described by the energy $E_2$ of Eq.~(\ref{eq:ham_2}). The rotational degrees of freedom are taken into account by including both the angles $\theta_i$ that the magnetic dipoles form with the chain axis and the dipole azimuthal angles $\phi_i$ as independent variables. This means that the magnetic moments are coupled to elastic deformations through the energetic contributions of Eqs.~(\ref{eq:h_dip_0}), (\ref{eq:h_D}), and (\ref{eq:h_tau}). As in the previous section, only spatially homogeneous solutions are considered for the displacements $a_i = a$ and the angles $\theta_i = \theta \quad \forall \, i = 1, \cdots, N$.

 The energetic contribution in Eq.~(\ref{eq:h_D}) introduces the memory angle $\theta^{(0)}$ of the initial configuration, so that $E_D = 0$ for $\theta = \theta^{(0)}$. When modeling ferrogels, such a term mimics the effect of an imprinted orientation of the magnetic moments in the polymer matrix \cite{collin2003frozen,varga2003smart,gunther2012xray}. For example, an externally applied magnetic field can homogeneously orient the magnetic moments during the synthesis of the material. The memory of this state can be permanently stored in the system through the chemical crosslinking process \cite{frickel2011magneto,messing2011cobalt}. 
An analogous conclusion follows from Eq.~(\ref{eq:h_tau}) for the memory of the initial relative azimuthal angles $\phi_i^{(0)}-\phi_j^{(0)}$ between neighboring dipoles. 

In summary, it is assumed that the magnetic moments are homogeneously aligned by an external magnetic field during cross-linking, so that $\theta_i^{(0)}=\theta^{(0)}$ and $\phi_i^{(0)} - \phi_j^{(0)} = 0 \quad \forall \, i, j = 1, \cdots, N$. When the magnetic field is then switched off after crosslinking, the system relaxes to minimize its energy, given by $E_2 = E_1 + E_D + E_\tau$ in Eq.~(\ref{eq:ham_2}). 
Consequently the equilibrium state is characterized by the system of equations 
\begin{equation}\label{eq:sys}
 \left\{ 
  \begin{array}{l}
   \partial E_2 / \partial \phi_i = 0 \quad \forall i = 1, \cdots, N,\\
   \partial E_2 / \partial \theta = 0, \\
   \partial E_2 / \partial a = 0.
  \end{array} \right.
\end{equation}

\begin{figure}
\begin{center} 
\includegraphics[width=8.5cm]{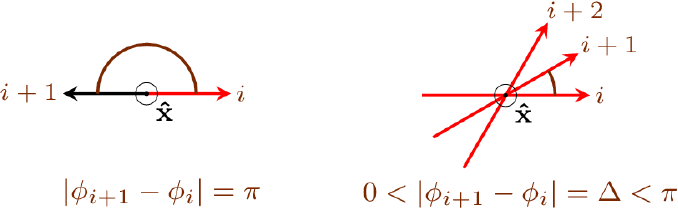}
\caption{(Color online) 
Antiferromagnetic (left) and spiral-like (right) configurations of the magnetic moments when seen from along the chain axis. In the antiferromagnetic case we plot two neighboring magnetic moments, in the spiral case we plot three neighboring magnetic moments. The antiferromagnetic case can be viewed as a degenerate spiral of $\Delta = \pi$.}
\label{fig0bc}
\end{center}
\end{figure}

 Using the ansatz $\phi_i = i \Delta \quad \forall \, i = 1, \cdots, N$ we guarantee that all of the equations $\partial E_2 / \partial \phi_i = 0$ are satisfied. 
In this way, we investigate the existence of solutions of spiral-like magnetization, where $\Delta$ is the relative azimuthal angle between the magnetic moments of any two neighboring particles (see Fig.~\ref{fig0bc}). As a result, the energy $E_2$ has to be minimized only with respect to the three remaining parameters $\Delta$, $\theta$, and $a$; i.e.\ the first line in Eq.~(\ref{eq:sys}) is replaced by $\partial E_2/\partial\Delta=0$.

Our ansatz contains two limiting cases: for $\Delta=0$ all the dipole moments point into the same direction, so that this state may be called ferromagnetic; for $\Delta=\pi$ the dipoles are arranged in an alternating fashion, which we refer to as an antiferromagnetic state in this context (see Fig.~\ref{fig0bc}) although the angle $\theta$ with the chain axis can be less than $\pi/2$. 

As a consequence, the energy per particle becomes
\begin{eqnarray}
  \lefteqn{\hspace{-.5cm}\frac{E_2}{N} (a, \theta, \Delta) =}\nonumber\\ 
& &\frac{k}{2} (a - L)^2 + \frac{\mu_0 \zeta (3) m^2}{4 \pi a^3} \left[ \frac{f(\Delta)}{\zeta (3)} \sin^2 \theta  - 2 \cos^2 \theta \right]  \nonumber  \\ && {}+ D (\cos \theta - \cos \theta^{(0)})^2 + \tau (\cos \Delta - 1)^2 + \frac{E_{\rm HS}}{N} ,
\label{E_2}
\end{eqnarray}
 where $f(\Delta) = \sum\limits_{n = 1}^{\infty} \frac{\cos (n \Delta)}{n^3}$. This energy reduces to $E_1$ when we fix $\theta = \theta^{(0)}$ and $\Delta = 0$.

\begin{figure} 
\begin{center} 
\includegraphics[width=8.5cm]{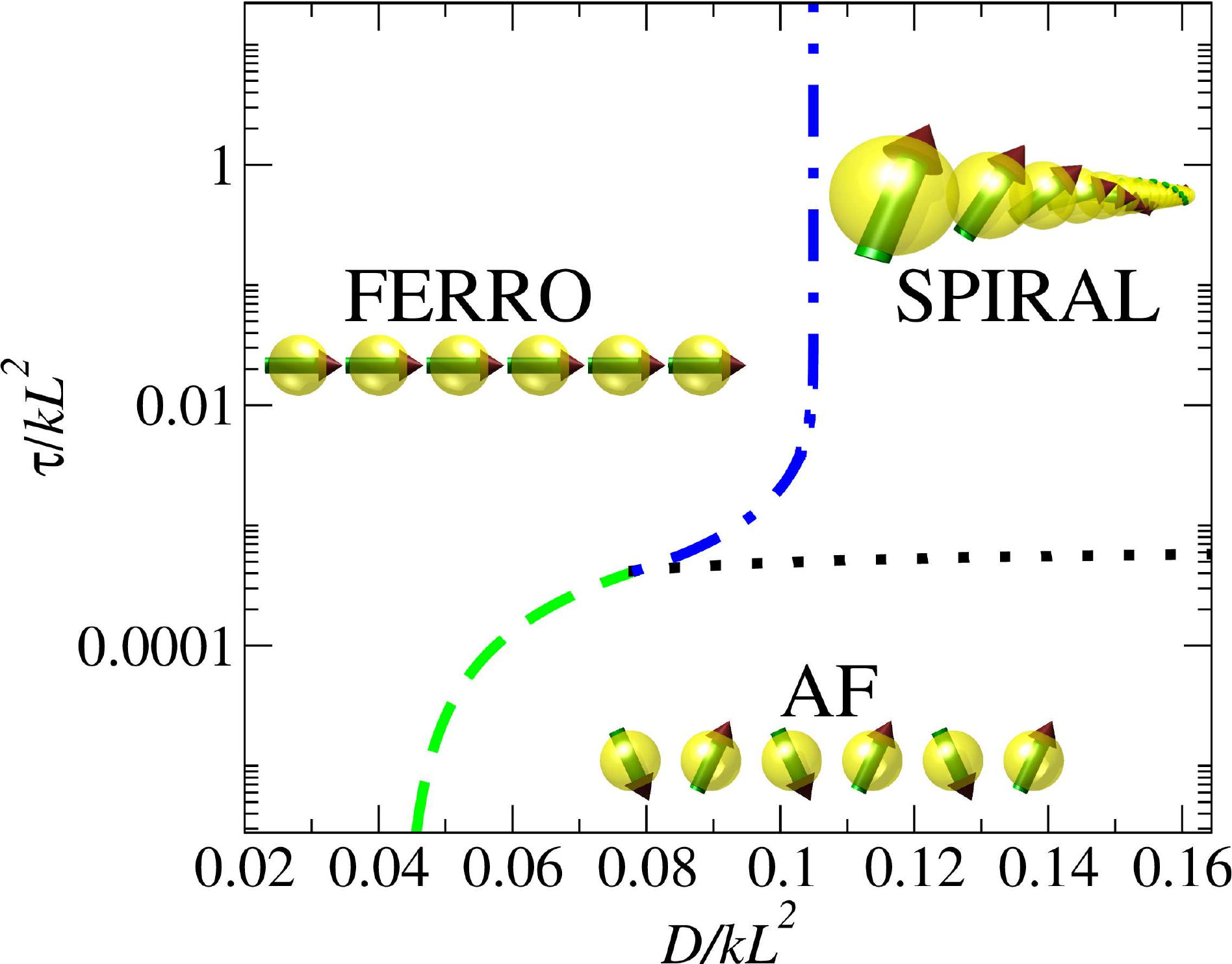}
\caption{(Color online) Phase diagram of the system for $m/m_0 \approx 0.3 $ and $\theta^{(0)}=\pi/4$ in the plane of the rescaled rotation parameters $D$ and $\tau$. We show the location of the three states ``FERRO'' (uniaxially ferromagnetic), ``SPIRAL'' (spirally magnetized), and ``AF'' (antiferromagnetic). 
The dash-dotted, dashed, and dotted lines correspond to the ``FERRO''--``SPIRAL'', ``FERRO''--``AF'', and ``SPIRAL''--``AF'' phase boundaries, respectively.
}
\label{fig_ph_dg}
\end{center}
\end{figure}

Searching for the minimum of the energy $E_2$, we identified three different cases for sufficiently high values of the magnetic moment $m/m_0$. A phase diagram in the plane of the rotational coupling parameters $(D,\tau)$ is reported for $m/m_0 \approx 0.3 $ in Fig.~\ref{fig_ph_dg}, where the three cases are illustrated and the boundaries between them are indicated. Here we set $\theta^{(0)}=\pi/4$. 

The first case is obtained for small values of $D$. Then rotations of the dipoles towards the chain axis do not lead to a significant energetic penalty. Therefore the dipolar energy is minimized by a ferromagnetic solution, in which all the magnetic moments are uniaxially aligned along the chain axis ($\theta = 0$). 

In the other two cases, rotations of the dipoles towards the chain axis cost a lot of energy, i.e.~$D$ is large. If, however, azimuthal rotations around the chain axis do not imply a significant energetic penalty (i.e.~$\tau$ is low), the dipolar energy is minimized by an alternating order. Here, the relative azimuthal angle between neighboring dipoles is $\Delta=\pi$, which we call the antiferromagnetic state. 

If, in the third case, also azimuthal rotations around the chain axis cost a lot of energy (i.e.~$\tau$ is large), the influence of the orientational memory is too strong for the limiting ferromagnetic or antiferromagnetic solutions to occur. As a compromise the system escapes into a spiral-like order of the magnetization directions. When the magnetic moment $m/m_0$ decreases, only spirally magnetized states are found below a certain threshold value. 

To obtain the phase boundaries in Fig.~\ref{fig_ph_dg}, we compared the energies of the three different states -- uniaxially ferromagnetic, antiferromagnetic, and spirally magnetized -- to each other. The phase boundaries were obtained and are drawn as equal-energy lines between all three pairs of different states.

During the remaining part of this section, we outline the impact of the orientational memory on the phase diagram in Fig.~\ref{fig3}. We recall that the external magnetic field is switched off when we are looking for the ground state of the system. It can therefore not be used to tune the magnitude of the magnetic moment $m/m_0$ as in the previous section. Consequently $m/m_0$ must rather be viewed as a material parameter in the following. 

\begin{figure} 
\begin{center} 
\includegraphics[width=8.5cm]{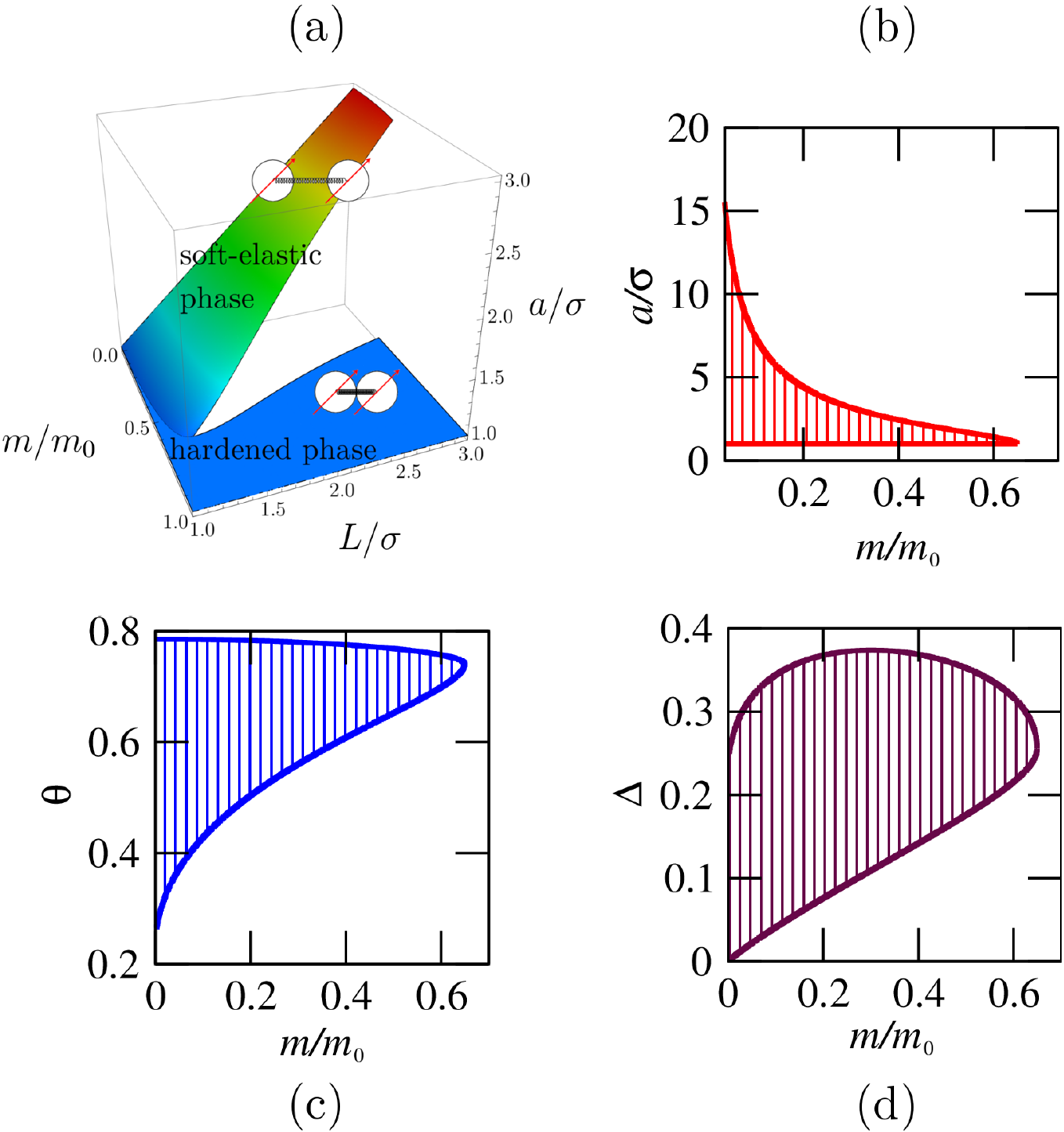}
\caption{(Color online) Phase diagram (a), as well as rescaled interparticle distance $a/\sigma$ (b), angle $\theta$ formed by the magnetic moments with the chain axis (c), and relative azimuthal angle $\Delta$ between neighboring magnetic moments (d), for states of coexistence. This coexistence is stressed by the tie lines. The data curves are obtained for a case of orientational memory that is characterized by the parameter values $\tau / kL^2 = 1$ and $D / kL^2 = 5$ (large-$\tau$ and large-$D$ regime). For the chosen parameters, the ground state of the magnetization is always spiral-like.}
\label{fig_ph_ori}
\end{center}
\end{figure}

We start with the situation of strong orientational memory, i.e.\ large values both for $D$ and $\tau$. In this case the ground state is reached by spirally magnetized states. 
Fig.~\ref{fig_ph_ori} shows the phase diagram of a corresponding system for the dimensionless parameters $\tau / kL^2 = 1$ and $D / kL^2 = 5$. 
As can be seen from Fig.~\ref{fig_ph_ori}~(a) the phase diagram is qualitatively similar to the non-memory case. Quantitatively, the orientational memory $\tau/kL^2 = 1$ and $D/kL^2 = 5$ leads to a decrease both in the critical magnetic moment $(m/m_0)_c$, from previously $0.76 $ to now $0.65 $, and in the critical quantity $(L/\sigma)_c$, from previously $5/4$ to now $1.22$. For the states of coexisting ``soft-elastic'' and ``hardened'' states we report in Fig.~\ref{fig_ph_ori} (b)--(d) the rescaled interparticle distance $a/\sigma(m/m_0)$, the angle $\theta(m/m_0)$ formed by the magnetic moments with the chain axis, and the relative azimuthal angle $\Delta(m/m_0)$ between neighboring magnetic moments. At variance with the non-memory case, for each $m/m_0 < (m/m_0)_c$ the two values of $a$ at coexistence correspond to two non-trivial values of $\theta$ and $\Delta$ different from $0$ and $\pi$, as shown in Fig.~\ref{fig_ph_ori}~(c) and (d). Below we illustrate how these curves change when $\tau$ and $D$ are varied.

\begin{figure} 
\begin{center} 
\includegraphics[width=8.5cm]{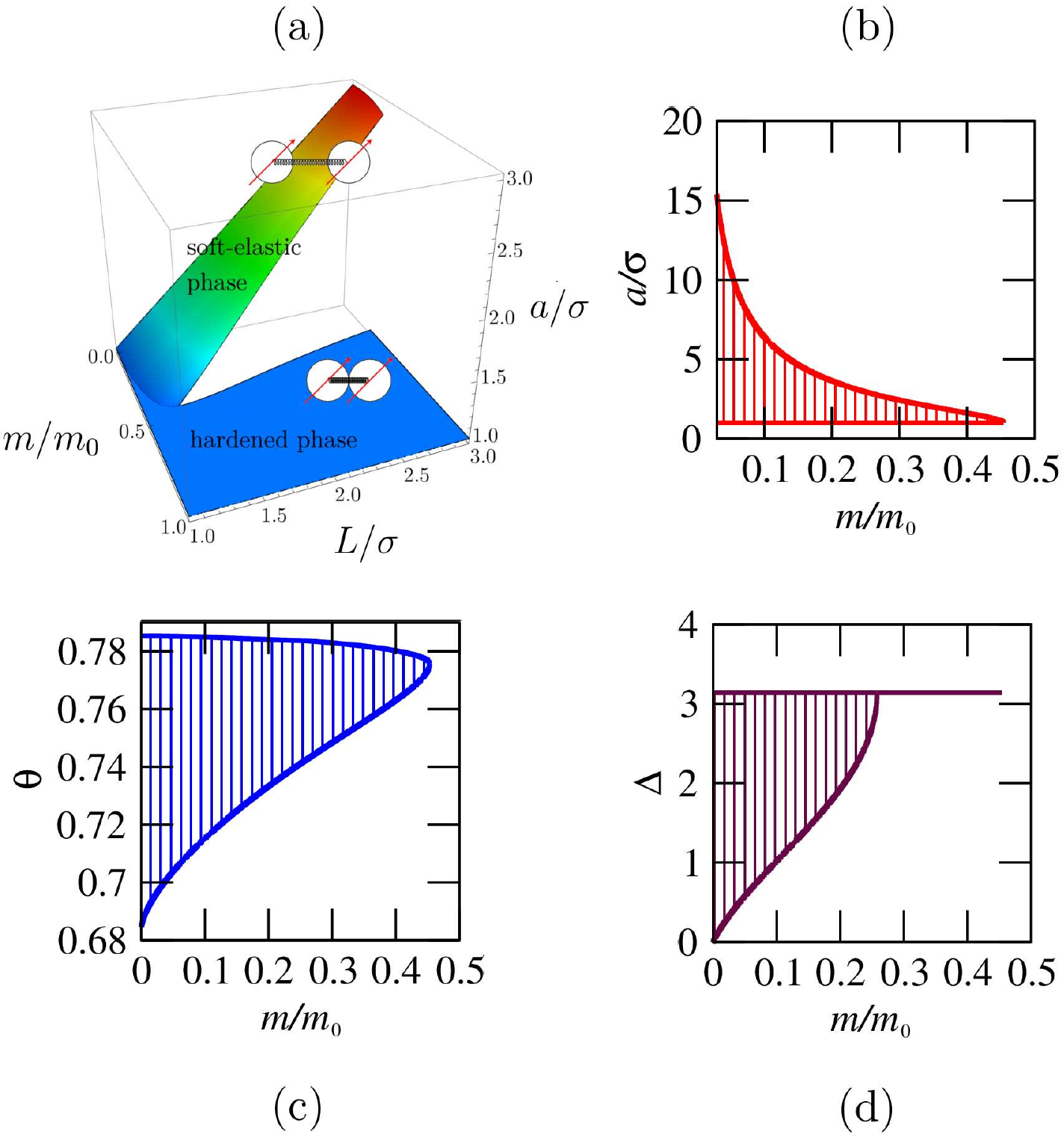}
\caption{(Color online) Phase diagram (a), as well as rescaled interparticle distance $a/\sigma$ (b), angle $\theta$ formed by the magnetic moments with the chain axis (c), and relative azimuthal angle $\Delta$ between neighboring magnetic moments (d), for states of coexistence. This coexistence is stressed by the tie lines. The data curves are obtained for a case of orientational memory that is characterized by the parameter values $\tau / kL^2 = 5 \cdot 10^{-4}$ and $D / kL^2 = 5$ (small-$\tau$ and large-$D$ regime). The ground state of the magnetization is spiral-like for $m/m_0 \lesssim 0.25 $, for larger values of $m/m_0$ it becomes antiferromagnetic.}
\label{fig_ph_alt}
\end{center}
\end{figure}

Next, we consider the case of low orientational memory concerning azimuthal rotations of the dipoles around the chain axis, but strong orientational memory for rotations towards the chain axis. In other words, $\tau$ is small and $D$ is large. 
 As an example, the phase behavior of the system is reported for $\tau / kL^2 = 5 \cdot 10^{-4}$ and $D/kL^2 = 5$ in Fig.~\ref{fig_ph_alt}. 
\begin{figure} 
\begin{center} 
\includegraphics[width=8.5cm]{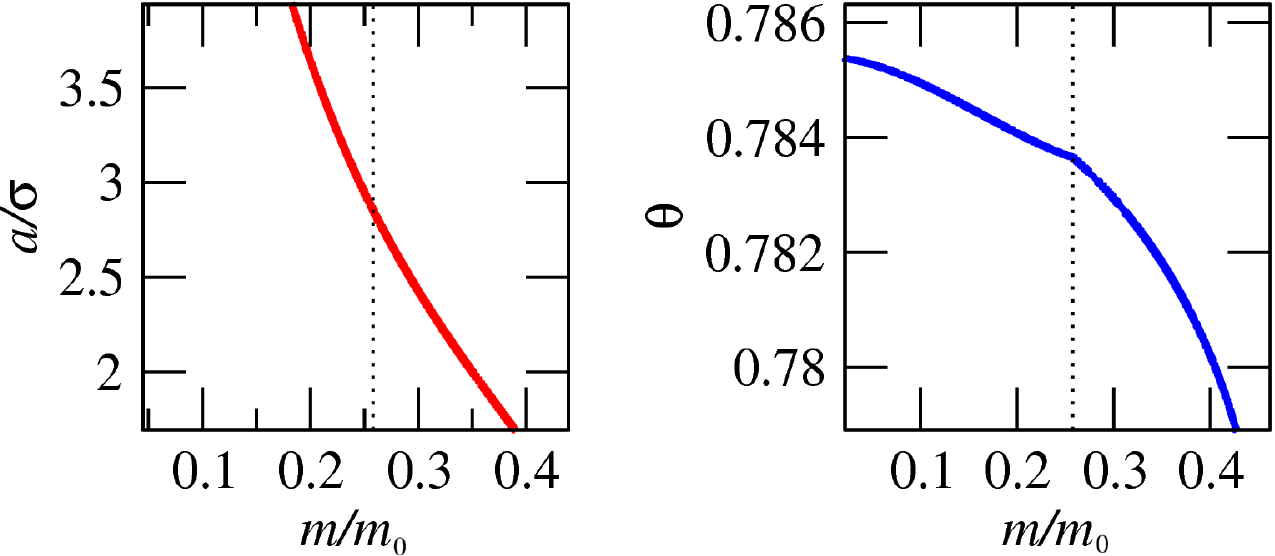}
\caption{(Color online) Magnification of the upper branches of the $a/\sigma$ and $\theta$ coexistence curves in Fig.~\ref{fig_ph_alt} (b) and (c). The characteristics of a second-order phase transition are visible in the $a/\sigma$ and $\theta$ variables at the point above which only antiferromagnetic states are found ($m/m_0 \approx 0.25)$. In the curve for $\theta$ the discontinuity in the first derivative is more evident.}
\label{fig_ph_alt_2}
\end{center}
\end{figure}
The $a/\sigma(m/m_0)$ and $\theta(m/m_0)$ coexistence curves are qualitatively similar to the previous large-$\tau$ and large-$D$ case. Now, the critical magnetic moment $(m/m_0)_c$ is even smaller ($(m/m_0)_c \approx 0.45 $) while the critical quantity $(L/\sigma)_c$ is approximately the same as in the non-memory case. The $\Delta (m/m_0)$ coexistence curve instead is  qualitatively different when we compare Figs.~\ref{fig_ph_ori}~(d) and \ref{fig_ph_alt}~(d). As can be seen from Fig.~\ref{fig_ph_alt}~(d), for $m/m_0 \lesssim 0.25 $ we find a spirally magnetized state of $0<\Delta<\pi$ which coexists with the antiferromagnetic state $\Delta = \pi$, while for $m/m_0 \gtrsim 0.25 $ the ground state is given by the antiferromagnetic solution.
It is interesting to note that at the value of the magnetic moment above which only antiferromagnetic states are found, $m/m_0 \approx 0.25 $, both upper branches of the $a/\sigma (m/m_0)$ and $\theta(m/m_0)$ coexistence curves are continuous but their first derivatives are discontinuous. These are the characteristics of a further, second-order phase transition in the variables $a/\sigma$ and $\theta$. Fig.~\ref{fig_ph_alt_2} stresses these additional results.

\begin{figure} 
\begin{center} 
\includegraphics[width=8.5cm]{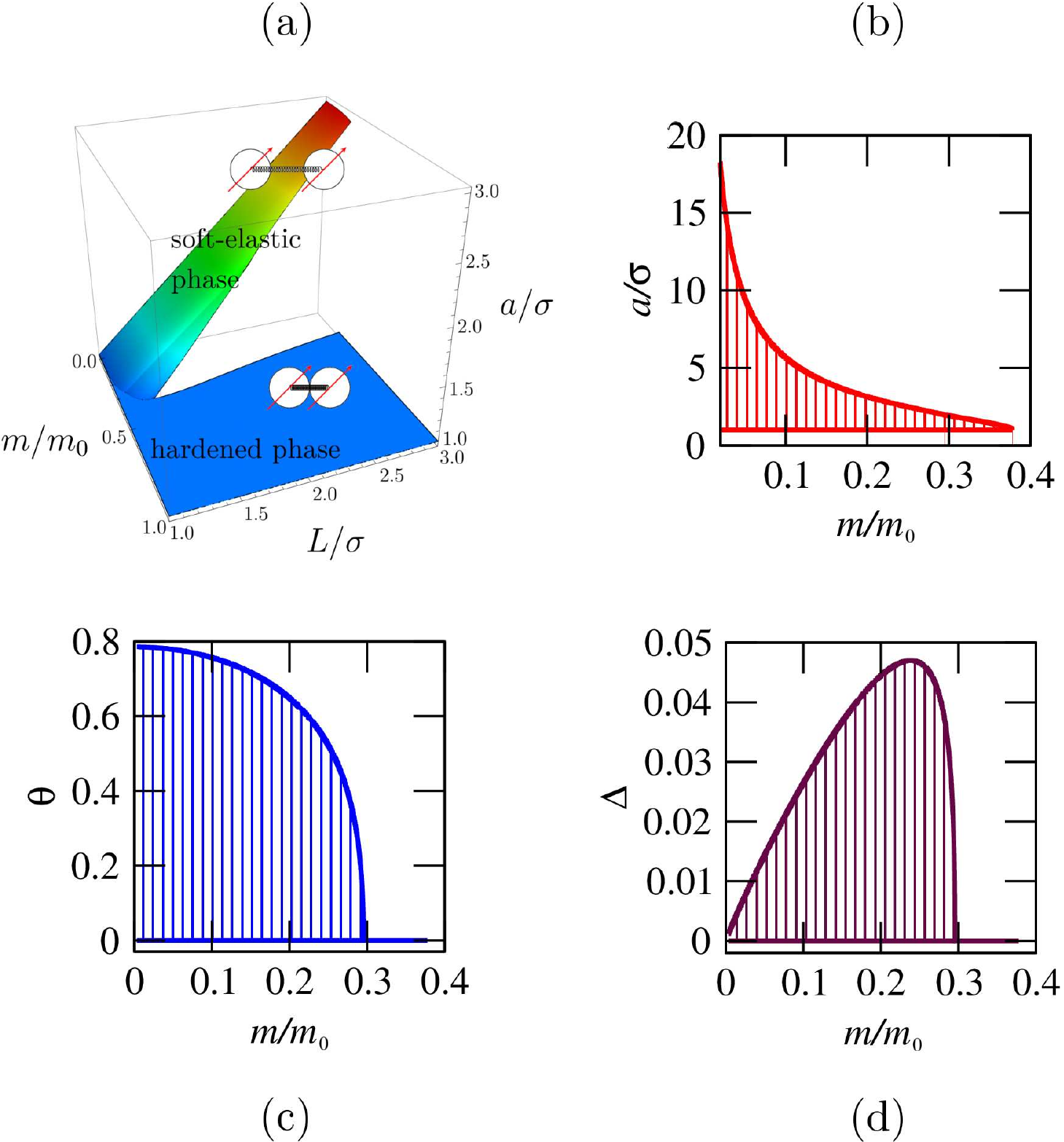}
\caption{(Color online) Phase diagram (a), as well as rescaled interparticle distance $a/\sigma$ (b), angle $\theta$ formed by the magnetic moments with the chain axis (c), and relative azimuthal angle $\Delta$ between neighboring magnetic moments (d), for states of coexistence. This coexistence is stressed by the tie lines. The data curves are obtained for a case of orientational memory that is characterized by the parameter values $\tau / kL^2 = 2.5$ and $D / kL^2 = 0.105$ (large-$\tau$ and small-$D$ regime). The ground state of the magnetization is spiral-like for $m/m_0 \lesssim 0.3 $, for larger values of $m/m_0$ it becomes ferromagnetic.}
\label{fig_ph_ferro}
\end{center}
\end{figure}

\begin{figure} 
\begin{center} 
\includegraphics[width=8.5cm]{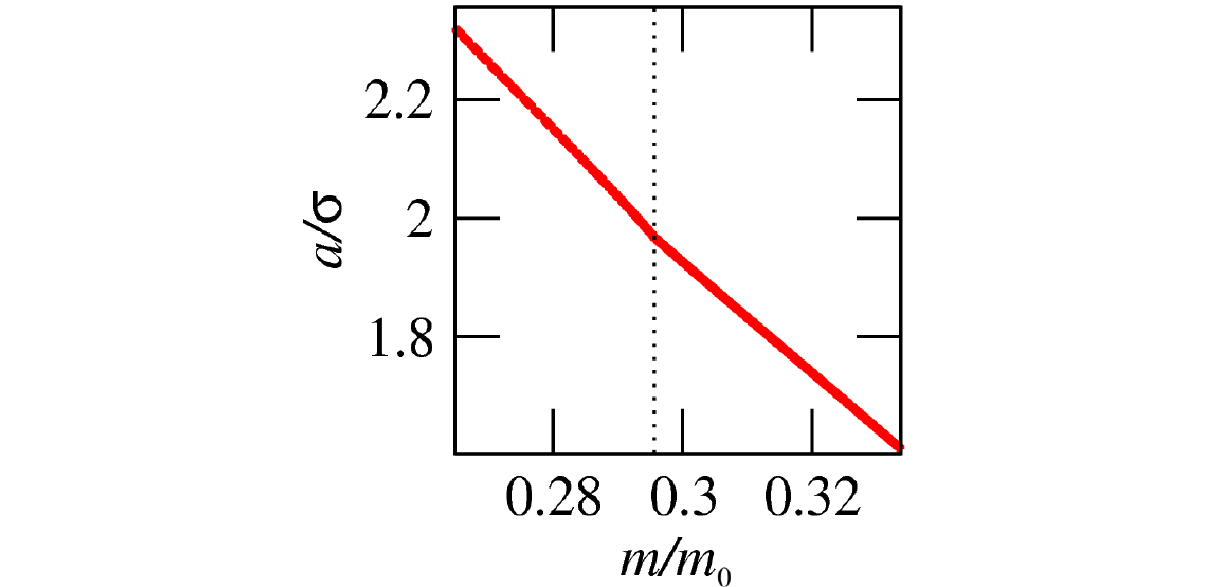}
\caption{(Color online) Magnification of the upper branch of the $a/\sigma$ coexistence curve in Fig.~\ref{fig_ph_ferro} (b). The characteristics of a second-order phase transition are visible in the $a/\sigma$ variable at the point above which only uniaxial ferromagnetic states are found ($m/m_0 \approx 0.3)$.}
\label{fig_ph_ferro_2}
\end{center}
\end{figure}

 In Fig.~\ref{fig_ph_ferro} the phase diagram and the coexistence curves are reported for the dimensionless parameters $\tau / kL^2 = 2.5$ and $D/kL^2 = 0.105$, which can be considered as a large value for $\tau$ but as a small value for $D$. 
In this case the ground state is reached for $m/m_0 \lesssim 0.3 $ by a spirally magnetized state which coexists with the ferromagnetic state $\theta = 0$. For $m/m_0 \gtrsim 0.3 $ the ground state is given by the ferromagnetic state. Again the characteristics of a second-order transition appear in the variable $a/\sigma$, now at the value of the magnetic moment above which only uniaxial ferromagnetic states are found, i.e.~$m/m_0 \approx 0.3$. The upper branch of the $a/\sigma (m/m_0)$ coexisting curve is continuous but its first derivative is discontinuous, as stressed by Fig.~\ref{fig_ph_ferro_2}. 
Furthermore, in the ferromagnetic state the orientational memory effects are outbalanced. The solution $\theta = 0$ makes the memory terms vanish in the expression for the energy of the system. Consequently the critical point corresponds to the one without orientational memory, but in the presence of an external magnetic field aligned with the system axis, $\theta_B = 0$. Therefore $(L/\sigma)_c = 5/4$ and $(m/m_0)_c \approx 0.3778 $ are obtained as from Eqs.~(\ref{coordinates_cpL})--(\ref{coordinates_cpa}).

\section{Conclusions} \label{sec5}

In this paper we introduced a simplified model for quasi one-dimensional ferrogels. We mapped the behavior of dipolar magnetic particles that are positionally and orientationally coupled to elastic deformations of the surrounding polymer matrix.
The particles are represented by hard spheres and connected to each other by identical harmonic springs of a given rest length; furthermore, we take into account the dipole-dipole magnetic interparticle potential. This model concerns ferrogels at the mesoscopic level and incorporates the magnetic particles explicitly, with an approach similar to Ivaneyko et al.~\cite{ivaneyko2011magneto}, but it is the first model to introduce explicitly the rotational coupling between the dipole orientations and the polymer matrix. 

 Starting from dipole orientations imprinted into the materials during the crosslinking process \cite{collin2003frozen,messing2011cobalt}, the model can be used for moderate dipolar rotation angles to mimic the wrapping of the polymer chains around the magnetic particles when polymer chains are covalently bound to the particle surfaces \cite{frickel2011magneto} and the magnetic particles rotate. This effect has recently been outlined in a more microscopic simulation approach \cite{weeber2012deformation}.

 We first studied the phase diagram of this model without orientational memory. This corresponds to the case of superparamagnetic particles, in which the direction of the magnetic moment can reorient with respect to the particle axes, or to the case of ferromagnetic particles that can rotate with respect to the polymer matrix without energetic penalty on the considered time scales. The orientation of the magnetic moments is determined by a strong external magnetic field. A hard-sphere repulsion between the particles stabilizes the system. For these conditions and neglecting thermal fluctuations, we investigated the behavior of the elastic modulus and the interparticle distance under the assumption of spatial homogeneity. 
The behavior of the system can change radically as a function of the angle that the strong external magnetic field forms with the particle chain. Depending on this angle, the dipolar interaction can be attractive or repulsive. In the attractive case we find a phase transition between ``hardened'' states on the one hand, where the particles touch each other and the compressive elastic modulus diverges, and ``soft-elastic'' states on the other hand, characterized by a non-zero interparticle distance and a finite elastic modulus. The phase transition is continuous for small rescaled initial interparticle distances and discontinuous for larger rescaled initial interparticle distances. A critical point corresponding to a second-order phase transition is identified, where the susceptibility of magnetostriction diverges with the critical exponent $\beta = 1/2$ as a function of the magnetic moment. The elastic modulus is a decreasing function of the magnetic moment and goes to zero at the critical point. When the dipolar interactions are repulsive, the elastic modulus and the equilibrium interparticle distance are increasing functions of the magnetic moment. This case is not as rich because there is no phase transition.

 An additional orientational memory that explicitly couples the orientations of the magnetic moments to the elastic deformations changes both the coexistence curves and the location of the critical point. We investigated, in the absence of an externally applied magnetic field, the case of a memory angle such that the dipole-dipole interaction is attractive. Similarly to the non-memory case, the phase diagram shows ``hardened'' and ``soft-elastic'' states, but there are quantitative differences: both the critical magnetic moment and the critical rescaled initial interparticle distance are smaller. Assuming initially homogeneously oriented dipolar moments along the chain, we asked the question, to which ground state the system relaxes as a function of the orientational memory of the initial state. Depending on the strength and nature of the orientational memory and the magneto-mechanical coupling, we find that the system relaxes to a uniaxial ferromagnetic, an antiferromagnetic, or a spirally magnetized state. 
 
In summary, we find that the nature of the orientational memory and of the magneto-mechanical coupling significantly determines the ground state of the system. Depending on the strength of the magneto-mechanical coupling, the magnetic moments can more or less easily rotate relatively to the surrounding elastic polymer network. From a macroscopic point of view, the importance of such relative rotations for the system behavior has been outlined before for polymeric liquid crystalline materials \cite{degennes1980, brand1994electrohydrodynamics}. A corresponding nonlinear macroscopic description has been established \cite{menzel2007nonlinear, menzel2009nonlinear, menzel2009response}. Furthermore, the relevance of relative rotations has also been discussed for a macroscopic characterization of ferrogels \cite{bohlius2004macroscopic, brand2011physical}. One of our future tasks will be to derive the connection between a mesoscopic approach as the one introduced in the present paper and the macroscopic characterizations. 

We stress that our results rely on a simplified model: first, every particle is represented by a hard sphere, whereas it is known already for non-crosslinked systems that an elongation of the particles can influence the structural order in dipolar systems \cite{miller2009dynamical}. To mimic the presence of the polymer network, springs are attached to the particle centers, which clearly neglects the complexity of attaching the springs on the surfaces or considering the wrapping of the springs around the surfaces for large rotation angles \cite{weeber2012deformation}. The latter effect is modeled by the orientational memory terms. 
Moreover, we used harmonic springs to mimic the elastic behavior. In particular when the magnetic particles come very close, however, nonlinear elasticity effects can become important. They may to some degree mask the predicted behavior, for example the strong pre-transitional reduction of the compressive elastic modulus preceding the phase transition from ``soft-elastic'' to ``hardened'' states. 

Furthermore, we neglected thermal fluctuations of the magnetic particles, i.e.\ we treated the particles as if they were at zero temperature. For this case, we found the phase transition between ``hardened'' and ``soft-elastic'' states in one dimension. Generally, at finite temperatures and in one dimension, real thermodynamic phase transitions only exist if the interaction forces do not decrease too quickly with separation distance \cite{dobrushin1973analyticity,frohlich1982phase}. The dipolar interactions in our case do not satisfy this condition. However, smoothened transitions should still be observable for chains of sufficiently finite length. Apart from that, our coupling constants can become temperature-dependent. As a simple example, the fixed-junction model for ideal rubber elasticity suggests that the elastic restoring forces should increase linearly with temperature \cite{strobl1997physics}. This translates into a corresponding dependence of our harmonic spring constant.

Finally, our model is one-dimensional and confines the spatial distribution of the magnetic particles to one line. Hence it is not able to cover important features such as the interplay of dipolar attraction and repulsion along different directions. For example, a two-dimensional array of aligned magnetic dipole particles has attractive dipolar interactions along the direction of the magnetic moments, while repulsive interactions occur in the orthogonal direction \cite{ivaneyko2011magneto}. 
It has been demonstrated theoretically, that different higher-dimensional spatial arrangements  can qualitatively change the characteristics of the system \cite{ivaneyko2012effects,han2013field}. We consider the behavior of the elastic modulus for deformations along an externally induced particle magnetization. A net increase of the elastic modulus with increasing magnetization is mostly observed for real materials, in contrast to our case. It has been demonstrated that a more isotropic particle distribution of hexagonal symmetry in three spatial dimensions reproduces the effect \cite{ivaneyko2012effects}. The latter results from the interplay of attractive and repulsive forces along different spatial directions.

Nevertheless, the presented approach can be employed to study effectively one-dimensional arrangements, for example when flexible supramolecular magnetic filaments \cite{sanchez2013filaments,cerda2013filaments} are fixed along a line. Our model explicitly introduces the memory of the initial magnetic orientations. We believe it to be a starting point to investigate the physics of ferrogel systems when extending it to more spatial dimensions and including thermal fluctuations.

\begin{acknowledgments}

The authors thank H.~R.~Brand and G.~K.~Auernhammer for helpful discussions and gratefully acknowledge the new environment and support from the recently founded SPP 1681 by the Deutsche Forschungsgemeinschaft.

\end{acknowledgments}


\begin{thebibliography}{67}%
\makeatletter
\providecommand \@ifxundefined [1]{%
 \@ifx{#1\undefined}
}%
\providecommand \@ifnum [1]{%
 \ifnum #1\expandafter \@firstoftwo
 \else \expandafter \@secondoftwo
 \fi
}%
\providecommand \@ifx [1]{%
 \ifx #1\expandafter \@firstoftwo
 \else \expandafter \@secondoftwo
 \fi
}%
\providecommand \natexlab [1]{#1}%
\providecommand \enquote  [1]{``#1''}%
\providecommand \bibnamefont  [1]{#1}%
\providecommand \bibfnamefont [1]{#1}%
\providecommand \citenamefont [1]{#1}%
\providecommand \href@noop [0]{\@secondoftwo}%
\providecommand \href [0]{\begingroup \@sanitize@url \@href}%
\providecommand \@href[1]{\@@startlink{#1}\@@href}%
\providecommand \@@href[1]{\endgroup#1\@@endlink}%
\providecommand \@sanitize@url [0]{\catcode `\\12\catcode `\$12\catcode
  `\&12\catcode `\#12\catcode `\^12\catcode `\_12\catcode `\%12\relax}%
\providecommand \@@startlink[1]{}%
\providecommand \@@endlink[0]{}%
\providecommand \url  [0]{\begingroup\@sanitize@url \@url }%
\providecommand \@url [1]{\endgroup\@href {#1}{\urlprefix }}%
\providecommand \urlprefix  [0]{URL }%
\providecommand \Eprint [0]{\href }%
\providecommand \doibase [0]{http://dx.doi.org/}%
\providecommand \selectlanguage [0]{\@gobble}%
\providecommand \bibinfo  [0]{\@secondoftwo}%
\providecommand \bibfield  [0]{\@secondoftwo}%
\providecommand \translation [1]{[#1]}%
\providecommand \BibitemOpen [0]{}%
\providecommand \bibitemStop [0]{}%
\providecommand \bibitemNoStop [0]{.\EOS\space}%
\providecommand \EOS [0]{\spacefactor3000\relax}%
\providecommand \BibitemShut  [1]{\csname bibitem#1\endcsname}%
\let\auto@bib@innerbib\@empty
\bibitem [{\citenamefont {Filipcsei}\ \emph {et~al.}(2007)\citenamefont
  {Filipcsei}, \citenamefont {Csetneki}, \citenamefont {Szil{\'a}gyi},\ and\
  \citenamefont {Zr\'{i}nyi}}]{filipcsei2007magnetic}%
  \BibitemOpen
  \bibfield  {author} {\bibinfo {author} {\bibfnamefont {G.}~\bibnamefont
  {Filipcsei}}, \bibinfo {author} {\bibfnamefont {I.}~\bibnamefont {Csetneki}},
  \bibinfo {author} {\bibfnamefont {A.}~\bibnamefont {Szil{\'a}gyi}}, \ and\
  \bibinfo {author} {\bibfnamefont {M.}~\bibnamefont {Zr\'{i}nyi}},\
  }\href@noop {} {\bibfield  {journal} {\bibinfo  {journal} {Adv. Polym. Sci.}\
  }\textbf {\bibinfo {volume} {206}},\ \bibinfo {pages} {137} (\bibinfo {year}
  {2007})}\BibitemShut {NoStop}%
\bibitem [{\citenamefont {Deng}\ \emph {et~al.}(2006)\citenamefont {Deng},
  \citenamefont {Gong},\ and\ \citenamefont {Wang}}]{deng2006development}%
  \BibitemOpen
  \bibfield  {author} {\bibinfo {author} {\bibfnamefont {H.-X.}\ \bibnamefont
  {Deng}}, \bibinfo {author} {\bibfnamefont {X.-L.}\ \bibnamefont {Gong}}, \
  and\ \bibinfo {author} {\bibfnamefont {L.-H.}\ \bibnamefont {Wang}},\
  }\href@noop {} {\bibfield  {journal} {\bibinfo  {journal} {Smart Mater.
  Struct.}\ }\textbf {\bibinfo {volume} {15}},\ \bibinfo {pages} {N111}
  (\bibinfo {year} {2006})}\BibitemShut {NoStop}%
\bibitem [{\citenamefont {Stepanov}\ \emph {et~al.}(2007)\citenamefont
  {Stepanov}, \citenamefont {Abramchuk}, \citenamefont {Grishin}, \citenamefont
  {Nikitin}, \citenamefont {Kramarenko},\ and\ \citenamefont
  {Khokhlov}}]{stepanov2007effect}%
  \BibitemOpen
  \bibfield  {author} {\bibinfo {author} {\bibfnamefont {G.~V.}\ \bibnamefont
  {Stepanov}}, \bibinfo {author} {\bibfnamefont {S.~S.}\ \bibnamefont
  {Abramchuk}}, \bibinfo {author} {\bibfnamefont {D.~A.}\ \bibnamefont
  {Grishin}}, \bibinfo {author} {\bibfnamefont {L.~V.}\ \bibnamefont
  {Nikitin}}, \bibinfo {author} {\bibfnamefont {E.~Y.}\ \bibnamefont
  {Kramarenko}}, \ and\ \bibinfo {author} {\bibfnamefont {A.~R.}\ \bibnamefont
  {Khokhlov}},\ }\href@noop {} {\bibfield  {journal} {\bibinfo  {journal}
  {Polymer}\ }\textbf {\bibinfo {volume} {48}},\ \bibinfo {pages} {488}
  (\bibinfo {year} {2007})}\BibitemShut {NoStop}%
\bibitem [{\citenamefont {Chen}\ \emph {et~al.}(2007)\citenamefont {Chen},
  \citenamefont {Gong}, \citenamefont {Jiang}, \citenamefont {Yao},
  \citenamefont {Deng},\ and\ \citenamefont {Li}}]{chen2007investigation}%
  \BibitemOpen
  \bibfield  {author} {\bibinfo {author} {\bibfnamefont {L.}~\bibnamefont
  {Chen}}, \bibinfo {author} {\bibfnamefont {X.-L.}\ \bibnamefont {Gong}},
  \bibinfo {author} {\bibfnamefont {W.-Q.}\ \bibnamefont {Jiang}}, \bibinfo
  {author} {\bibfnamefont {J.-J.}\ \bibnamefont {Yao}}, \bibinfo {author}
  {\bibfnamefont {H.-X.}\ \bibnamefont {Deng}}, \ and\ \bibinfo {author}
  {\bibfnamefont {W.-H.}\ \bibnamefont {Li}},\ }\href@noop {} {\bibfield
  {journal} {\bibinfo  {journal} {J. Mater. Sci.}\ }\textbf {\bibinfo {volume}
  {42}},\ \bibinfo {pages} {5483} (\bibinfo {year} {2007})}\BibitemShut
  {NoStop}%
\bibitem [{\citenamefont {B{\"o}se}\ and\ \citenamefont
  {R{\"o}der}(2009)}]{bose2009magnetorheological}%
  \BibitemOpen
  \bibfield  {author} {\bibinfo {author} {\bibfnamefont {H.}~\bibnamefont
  {B{\"o}se}}\ and\ \bibinfo {author} {\bibfnamefont {R.}~\bibnamefont
  {R{\"o}der}},\ }\href@noop {} {\bibfield  {journal} {\bibinfo  {journal} {J.
  Phys.: Conf. Ser.}\ }\textbf {\bibinfo {volume} {149}},\ \bibinfo {pages}
  {012090} (\bibinfo {year} {2009})}\BibitemShut {NoStop}%
\bibitem [{\citenamefont {Sun}\ \emph {et~al.}(2008)\citenamefont {Sun},
  \citenamefont {Gong}, \citenamefont {Jiang}, \citenamefont {Li},
  \citenamefont {Xu},\ and\ \citenamefont {Li}}]{sun2008study}%
  \BibitemOpen
  \bibfield  {author} {\bibinfo {author} {\bibfnamefont {T.~L.}\ \bibnamefont
  {Sun}}, \bibinfo {author} {\bibfnamefont {X.~L.}\ \bibnamefont {Gong}},
  \bibinfo {author} {\bibfnamefont {W.~Q.}\ \bibnamefont {Jiang}}, \bibinfo
  {author} {\bibfnamefont {J.~F.}\ \bibnamefont {Li}}, \bibinfo {author}
  {\bibfnamefont {Z.~B.}\ \bibnamefont {Xu}}, \ and\ \bibinfo {author}
  {\bibfnamefont {W.}~\bibnamefont {Li}},\ }\href@noop {} {\bibfield  {journal}
  {\bibinfo  {journal} {Polym. Test.}\ }\textbf {\bibinfo {volume} {27}},\
  \bibinfo {pages} {520} (\bibinfo {year} {2008})}\BibitemShut {NoStop}%
\bibitem [{\citenamefont
  {Rosensweig}(1985)}]{rosensweig1985ferrohydrodynamics}%
  \BibitemOpen
  \bibfield  {author} {\bibinfo {author} {\bibfnamefont {R.~E.}\ \bibnamefont
  {Rosensweig}},\ }\href@noop {} {{\bibinfo {title}
  \textit{Ferrohydrodynamics}}}\ (\bibinfo  {publisher} {Cambridge University Press,
  Cambridge},\ \bibinfo {year} {1985})\BibitemShut {NoStop}%
\bibitem [{\citenamefont
  {Odenbach}(2003{\natexlab{a}})}]{odenbach2003ferrofluids}%
  \BibitemOpen
  \bibfield  {author} {\bibinfo {author} {\bibfnamefont {S.}~\bibnamefont
  {Odenbach}},\ }\href@noop {} {\bibfield  {journal} {\bibinfo  {journal}
  {Colloid Surface A}\ }\textbf {\bibinfo {volume} {217}},\ \bibinfo {pages}
  {171} (\bibinfo {year} {2003}{\natexlab{a}})}\BibitemShut {NoStop}%
\bibitem [{\citenamefont
  {Odenbach}(2003{\natexlab{b}})}]{odenbach2003magnetoviscous}%
  \BibitemOpen
  \bibfield  {author} {\bibinfo {author} {\bibfnamefont {S.}~\bibnamefont
  {Odenbach}},\ }\href@noop {} { {\bibinfo {title} \textit{Magnetoviscous effects
  in ferrofluids}}}\ (\bibinfo  {publisher} {Springer Berlin / Heidelberg},\
  \bibinfo {year} {2003})\BibitemShut {NoStop}%
\bibitem [{\citenamefont {Huke}\ and\ \citenamefont
  {L{\"u}cke}(2004)}]{huke2004magnetic}%
  \BibitemOpen
  \bibfield  {author} {\bibinfo {author} {\bibfnamefont {B.}~\bibnamefont
  {Huke}}\ and\ \bibinfo {author} {\bibfnamefont {M.}~\bibnamefont
  {L{\"u}cke}},\ }\href@noop {} {\bibfield  {journal} {\bibinfo  {journal}
  {Rep. Prog. Phys.}\ }\textbf {\bibinfo {volume} {67}},\ \bibinfo {pages}
  {1731} (\bibinfo {year} {2004})}\BibitemShut {NoStop}%
\bibitem [{\citenamefont {Odenbach}(2004)}]{odenbach2004recent}%
  \BibitemOpen
  \bibfield  {author} {\bibinfo {author} {\bibfnamefont {S.}~\bibnamefont
  {Odenbach}},\ }\href@noop {} {\bibfield  {journal} {\bibinfo  {journal} {J.
  Phys.: Condens. Matter}\ }\textbf {\bibinfo {volume} {16}},\ \bibinfo {pages}
  {R1135} (\bibinfo {year} {2004})}\BibitemShut {NoStop}%
\bibitem [{\citenamefont {Fischer}\ \emph {et~al.}(2005)\citenamefont
  {Fischer}, \citenamefont {Huke}, \citenamefont {L{\"u}cke},\ and\
  \citenamefont {Hempelmann}}]{fischer2005brownian}%
  \BibitemOpen
  \bibfield  {author} {\bibinfo {author} {\bibfnamefont {B.}~\bibnamefont
  {Fischer}}, \bibinfo {author} {\bibfnamefont {B.}~\bibnamefont {Huke}},
  \bibinfo {author} {\bibfnamefont {M.}~\bibnamefont {L{\"u}cke}}, \ and\
  \bibinfo {author} {\bibfnamefont {R.}~\bibnamefont {Hempelmann}},\
  }\href@noop {} {\bibfield  {journal} {\bibinfo  {journal} {J. Magn. Magn.
  Mater.}\ }\textbf {\bibinfo {volume} {289}},\ \bibinfo {pages} {74} (\bibinfo
  {year} {2005})}\BibitemShut {NoStop}%
\bibitem [{\citenamefont {Ilg}\ \emph {et~al.}(2005)\citenamefont {Ilg},
  \citenamefont {Kr{\"o}ger},\ and\ \citenamefont {Hess}}]{ilg2005structure}%
  \BibitemOpen
  \bibfield  {author} {\bibinfo {author} {\bibfnamefont {P.}~\bibnamefont
  {Ilg}}, \bibinfo {author} {\bibfnamefont {M.}~\bibnamefont {Kr{\"o}ger}}, \
  and\ \bibinfo {author} {\bibfnamefont {S.}~\bibnamefont {Hess}},\ }\href@noop
  {} {\bibfield  {journal} {\bibinfo  {journal} {J. Magn. Magn. Mater.}\
  }\textbf {\bibinfo {volume} {289}},\ \bibinfo {pages} {325} (\bibinfo {year}
  {2005})}\BibitemShut {NoStop}%
\bibitem [{\citenamefont {Embs}\ \emph {et~al.}(2006)\citenamefont {Embs},
  \citenamefont {May}, \citenamefont {Wagner}, \citenamefont {Kityk},
  \citenamefont {Leschhorn},\ and\ \citenamefont
  {L{\"u}cke}}]{embs2006measuring}%
  \BibitemOpen
  \bibfield  {author} {\bibinfo {author} {\bibfnamefont {J.~P.}\ \bibnamefont
  {Embs}}, \bibinfo {author} {\bibfnamefont {S.}~\bibnamefont {May}}, \bibinfo
  {author} {\bibfnamefont {C.}~\bibnamefont {Wagner}}, \bibinfo {author}
  {\bibfnamefont {A.~V.}\ \bibnamefont {Kityk}}, \bibinfo {author}
  {\bibfnamefont {A.}~\bibnamefont {Leschhorn}}, \ and\ \bibinfo {author}
  {\bibfnamefont {M.}~\bibnamefont {L{\"u}cke}},\ }\href@noop {} {\bibfield
  {journal} {\bibinfo  {journal} {Phys. Rev. E}\ }\textbf {\bibinfo {volume}
  {73}},\ \bibinfo {pages} {036302} (\bibinfo {year} {2006})}\BibitemShut
  {NoStop}%
\bibitem [{\citenamefont {Ilg}\ \emph {et~al.}(2006)\citenamefont {Ilg},
  \citenamefont {Coquelle},\ and\ \citenamefont {Hess}}]{ilg2006structure}%
  \BibitemOpen
  \bibfield  {author} {\bibinfo {author} {\bibfnamefont {P.}~\bibnamefont
  {Ilg}}, \bibinfo {author} {\bibfnamefont {E.}~\bibnamefont {Coquelle}}, \
  and\ \bibinfo {author} {\bibfnamefont {S.}~\bibnamefont {Hess}},\ }\href@noop
  {} {\bibfield  {journal} {\bibinfo  {journal} {J. Phys.: Condens. Matter}\
  }\textbf {\bibinfo {volume} {18}},\ \bibinfo {pages} {S2757} (\bibinfo {year}
  {2006})}\BibitemShut {NoStop}%
\bibitem [{\citenamefont {Gollwitzer}\ \emph {et~al.}(2007)\citenamefont
  {Gollwitzer}, \citenamefont {Matthies}, \citenamefont {Richter},
  \citenamefont {Rehberg},\ and\ \citenamefont
  {Tobiska}}]{gollwitzer2007surface}%
  \BibitemOpen
  \bibfield  {author} {\bibinfo {author} {\bibfnamefont {C.}~\bibnamefont
  {Gollwitzer}}, \bibinfo {author} {\bibfnamefont {G.}~\bibnamefont
  {Matthies}}, \bibinfo {author} {\bibfnamefont {R.}~\bibnamefont {Richter}},
  \bibinfo {author} {\bibfnamefont {I.}~\bibnamefont {Rehberg}}, \ and\
  \bibinfo {author} {\bibfnamefont {L.}~\bibnamefont {Tobiska}},\ }\href@noop
  {} {\bibfield  {journal} {\bibinfo  {journal} {J. Fluid Mech.}\ }\textbf
  {\bibinfo {volume} {571}},\ \bibinfo {pages} {455} (\bibinfo {year}
  {2007})}\BibitemShut {NoStop}%
\bibitem [{\citenamefont {de~Vicente}\ \emph {et~al.}(2011)\citenamefont
  {de~Vicente}, \citenamefont {Klingenberg},\ and\ \citenamefont
  {Hidalgo-Alvarez}}]{vicente2011magnetorheological}%
  \BibitemOpen
  \bibfield  {author} {\bibinfo {author} {\bibfnamefont {J.}~\bibnamefont
  {de~Vicente}}, \bibinfo {author} {\bibfnamefont {D.~J.}\ \bibnamefont
  {Klingenberg}}, \ and\ \bibinfo {author} {\bibfnamefont {R.}~\bibnamefont
  {Hidalgo-Alvarez}},\ }\href@noop {} {\bibfield  {journal} {\bibinfo
  {journal} {Soft Matter}\ }\textbf {\bibinfo {volume} {7}},\ \bibinfo {pages}
  {3701} (\bibinfo {year} {2011})}\BibitemShut {NoStop}%
\bibitem [{\citenamefont {Guan}\ \emph {et~al.}(2008)\citenamefont {Guan},
  \citenamefont {Dong},\ and\ \citenamefont {Ou}}]{guan2008magnetostrictive}%
  \BibitemOpen
  \bibfield  {author} {\bibinfo {author} {\bibfnamefont {X.}~\bibnamefont
  {Guan}}, \bibinfo {author} {\bibfnamefont {X.}~\bibnamefont {Dong}}, \ and\
  \bibinfo {author} {\bibfnamefont {J.}~\bibnamefont {Ou}},\ }\href@noop {}
  {\bibfield  {journal} {\bibinfo  {journal} {J. Magn. Magn. Mater.}\ }\textbf
  {\bibinfo {volume} {320}},\ \bibinfo {pages} {158} (\bibinfo {year}
  {2008})}\BibitemShut {NoStop}%
\bibitem [{\citenamefont {Zr\'{i}nyi}\ \emph {et~al.}(1997)\citenamefont
  {Zr\'{i}nyi}, \citenamefont {Barsi}, \citenamefont {Szab{\'o}},\ and\
  \citenamefont {Kilian}}]{zrinyi1997direct}%
  \BibitemOpen
  \bibfield  {author} {\bibinfo {author} {\bibfnamefont {M.}~\bibnamefont
  {Zr\'{i}nyi}}, \bibinfo {author} {\bibfnamefont {L.}~\bibnamefont {Barsi}},
  \bibinfo {author} {\bibfnamefont {D.}~\bibnamefont {Szab{\'o}}}, \ and\
  \bibinfo {author} {\bibfnamefont {H.-G.}\ \bibnamefont {Kilian}},\
  }\href@noop {} {\bibfield  {journal} {\bibinfo  {journal} {J. Chem. Phys.}\
  }\textbf {\bibinfo {volume} {106}},\ \bibinfo {pages} {5685} (\bibinfo {year}
  {1997})}\BibitemShut {NoStop}%
\bibitem [{\citenamefont {Szab\'{o}}\ \emph {et~al.}(1998)\citenamefont
  {Szab\'{o}}, \citenamefont {Szeghy},\ and\ \citenamefont
  {Zr\'{i}nyi}}]{szabo1998shape}%
  \BibitemOpen
  \bibfield  {author} {\bibinfo {author} {\bibfnamefont {D.}~\bibnamefont
  {Szab\'{o}}}, \bibinfo {author} {\bibfnamefont {G.}~\bibnamefont {Szeghy}}, \
  and\ \bibinfo {author} {\bibfnamefont {M.}~\bibnamefont {Zr\'{i}nyi}},\
  }\href@noop {} {\bibfield  {journal} {\bibinfo  {journal} {Macromolecules}\
  }\textbf {\bibinfo {volume} {31}},\ \bibinfo {pages} {6541} (\bibinfo {year}
  {1998})}\BibitemShut {NoStop}%
\bibitem [{\citenamefont {Filipcsei}\ and\ \citenamefont
  {Zr{\'\i}nyi}(2010)}]{filipcsei2010magnetodeformation}%
  \BibitemOpen
  \bibfield  {author} {\bibinfo {author} {\bibfnamefont {G.}~\bibnamefont
  {Filipcsei}}\ and\ \bibinfo {author} {\bibfnamefont {M.}~\bibnamefont
  {Zr{\'\i}nyi}},\ }\href@noop {} {\bibfield  {journal} {\bibinfo  {journal}
  {J. Phys.: Condens. Matter}\ }\textbf {\bibinfo {volume} {22}},\ \bibinfo
  {pages} {276001} (\bibinfo {year} {2010})}\BibitemShut {NoStop}%
\bibitem [{\citenamefont {Nikitin}\ \emph {et~al.}(2004)\citenamefont
  {Nikitin}, \citenamefont {Stepanov}, \citenamefont {Mironova},\ and\
  \citenamefont {Gorbunov}}]{nikitin2004magnetodeformational}%
  \BibitemOpen
  \bibfield  {author} {\bibinfo {author} {\bibfnamefont {L.~V.}\ \bibnamefont
  {Nikitin}}, \bibinfo {author} {\bibfnamefont {G.~V.}\ \bibnamefont
  {Stepanov}}, \bibinfo {author} {\bibfnamefont {L.~S.}\ \bibnamefont
  {Mironova}}, \ and\ \bibinfo {author} {\bibfnamefont {A.~I.}\ \bibnamefont
  {Gorbunov}},\ }\href@noop {} {\bibfield  {journal} {\bibinfo  {journal} {J.
  Magn. Magn. Mater.}\ }\textbf {\bibinfo {volume} {272}},\ \bibinfo {pages}
  {2072} (\bibinfo {year} {2004})}\BibitemShut {NoStop}%
\bibitem [{\citenamefont {Snyder}\ \emph {et~al.}(2010)\citenamefont {Snyder},
  \citenamefont {Nguyen},\ and\ \citenamefont {Ramanujan}}]{snyder2010design}%
  \BibitemOpen
  \bibfield  {author} {\bibinfo {author} {\bibfnamefont {R.~L.}\ \bibnamefont
  {Snyder}}, \bibinfo {author} {\bibfnamefont {V.~Q.}\ \bibnamefont {Nguyen}},
  \ and\ \bibinfo {author} {\bibfnamefont {R.~V.}\ \bibnamefont {Ramanujan}},\
  }\href@noop {} {\bibfield  {journal} {\bibinfo  {journal} {Smart Mater.
  Struct.}\ }\textbf {\bibinfo {volume} {19}},\ \bibinfo {pages} {055017}
  (\bibinfo {year} {2010})}\BibitemShut {NoStop}%
\bibitem [{\citenamefont {Collin}\ \emph {et~al.}(2003)\citenamefont {Collin},
  \citenamefont {Auernhammer}, \citenamefont {Gavat}, \citenamefont
  {Martinoty},\ and\ \citenamefont {Brand}}]{collin2003frozen}%
  \BibitemOpen
  \bibfield  {author} {\bibinfo {author} {\bibfnamefont {D.}~\bibnamefont
  {Collin}}, \bibinfo {author} {\bibfnamefont {G.~K.}\ \bibnamefont
  {Auernhammer}}, \bibinfo {author} {\bibfnamefont {O.}~\bibnamefont {Gavat}},
  \bibinfo {author} {\bibfnamefont {P.}~\bibnamefont {Martinoty}}, \ and\
  \bibinfo {author} {\bibfnamefont {H.~R.}\ \bibnamefont {Brand}},\ }\href@noop
  {} {\bibfield  {journal} {\bibinfo  {journal} {Macromol. Rapid Commun.}\
  }\textbf {\bibinfo {volume} {24}},\ \bibinfo {pages} {737} (\bibinfo {year}
  {2003})}\BibitemShut {NoStop}%
\bibitem [{\citenamefont {Varga}\ \emph {et~al.}(2003)\citenamefont {Varga},
  \citenamefont {Feh\'{e}r}, \citenamefont {Filipcsei},\ and\ \citenamefont
  {Zr{\'\i}nyi}}]{varga2003smart}%
  \BibitemOpen
  \bibfield  {author} {\bibinfo {author} {\bibfnamefont {Z.}~\bibnamefont
  {Varga}}, \bibinfo {author} {\bibfnamefont {J.}~\bibnamefont {Feh\'{e}r}},
  \bibinfo {author} {\bibfnamefont {G.}~\bibnamefont {Filipcsei}}, \ and\
  \bibinfo {author} {\bibfnamefont {M.}~\bibnamefont {Zr{\'\i}nyi}},\
  }\bibfield  {booktitle} {\textit{\bibinfo {booktitle} {Macromol. Symp.}},\
  }\href@noop {} {\ \textbf {\bibinfo {volume} {200}},\ \bibinfo {pages} {93}
  (\bibinfo {year} {2003})}\BibitemShut {NoStop}%
\bibitem [{\citenamefont {G{\"u}nther}\ \emph {et~al.}(2012)\citenamefont
  {G{\"u}nther}, \citenamefont {Borin}, \citenamefont {G{\"u}nther},\ and\
  \citenamefont {Odenbach}}]{gunther2012xray}%
  \BibitemOpen
  \bibfield  {author} {\bibinfo {author} {\bibfnamefont {D.}~\bibnamefont
  {G{\"u}nther}}, \bibinfo {author} {\bibfnamefont {D.~Y.}\ \bibnamefont
  {Borin}}, \bibinfo {author} {\bibfnamefont {S.}~\bibnamefont {G{\"u}nther}},
  \ and\ \bibinfo {author} {\bibfnamefont {S.}~\bibnamefont {Odenbach}},\
  }\href@noop {} {\bibfield  {journal} {\bibinfo  {journal} {Smart Mater.
  Struct.}\ }\textbf {\bibinfo {volume} {21}},\ \bibinfo {pages} {015005}
  (\bibinfo {year} {2012})}\BibitemShut {NoStop}%
\bibitem [{\citenamefont {Zr{\'\i}nyi}\ \emph {et~al.}(1996)\citenamefont
  {Zr{\'\i}nyi}, \citenamefont {Barsi},\ and\ \citenamefont
  {B{\"u}ki}}]{zrinyi1996deformation}%
  \BibitemOpen
  \bibfield  {author} {\bibinfo {author} {\bibfnamefont {M.}~\bibnamefont
  {Zr{\'\i}nyi}}, \bibinfo {author} {\bibfnamefont {L.}~\bibnamefont {Barsi}},
  \ and\ \bibinfo {author} {\bibfnamefont {A.}~\bibnamefont {B{\"u}ki}},\
  }\href@noop {} {\bibfield  {journal} {\bibinfo  {journal} {J. Chem. Phys.}\
  }\textbf {\bibinfo {volume} {104}},\ \bibinfo {pages} {8750} (\bibinfo {year}
  {1996})}\BibitemShut {NoStop}%
\bibitem [{\citenamefont {Zrinyi}\ \emph {et~al.}(1997)\citenamefont {Zrinyi},
  \citenamefont {Barsi},\ and\ \citenamefont {B{\"u}ki}}]{zrinyi1997ferrogel}%
  \BibitemOpen
  \bibfield  {author} {\bibinfo {author} {\bibfnamefont {M.}~\bibnamefont
  {Zrinyi}}, \bibinfo {author} {\bibfnamefont {L.}~\bibnamefont {Barsi}}, \
  and\ \bibinfo {author} {\bibfnamefont {A.}~\bibnamefont {B{\"u}ki}},\
  }\href@noop {} {\bibfield  {journal} {\bibinfo  {journal} {Polym. Gels
  Netw.}\ }\textbf {\bibinfo {volume} {5}},\ \bibinfo {pages} {415} (\bibinfo
  {year} {1997})}\BibitemShut {NoStop}%
\bibitem [{\citenamefont {Krekhova}\ \emph {et~al.}(2010)\citenamefont
  {Krekhova}, \citenamefont {Lang}, \citenamefont {Richter},\ and\
  \citenamefont {Schmalz}}]{krekhova2010thermoreversible}%
  \BibitemOpen
  \bibfield  {author} {\bibinfo {author} {\bibfnamefont {M.}~\bibnamefont
  {Krekhova}}, \bibinfo {author} {\bibfnamefont {T.}~\bibnamefont {Lang}},
  \bibinfo {author} {\bibfnamefont {R.}~\bibnamefont {Richter}}, \ and\
  \bibinfo {author} {\bibfnamefont {H.}~\bibnamefont {Schmalz}},\ }\href@noop
  {} {\bibfield  {journal} {\bibinfo  {journal} {Langmuir}\ }\textbf {\bibinfo
  {volume} {26}},\ \bibinfo {pages} {19181} (\bibinfo {year}
  {2010})}\BibitemShut {NoStop}%
\bibitem [{\citenamefont {Bonini}\ \emph {et~al.}(2008)\citenamefont {Bonini},
  \citenamefont {Lenz}, \citenamefont {Falletta}, \citenamefont {Ridi},
  \citenamefont {Carretti}, \citenamefont {Fratini}, \citenamefont
  {Wiedenmann},\ and\ \citenamefont {Baglioni}}]{bonini2008acrylamide}%
  \BibitemOpen
  \bibfield  {author} {\bibinfo {author} {\bibfnamefont {M.}~\bibnamefont
  {Bonini}}, \bibinfo {author} {\bibfnamefont {S.}~\bibnamefont {Lenz}},
  \bibinfo {author} {\bibfnamefont {E.}~\bibnamefont {Falletta}}, \bibinfo
  {author} {\bibfnamefont {F.}~\bibnamefont {Ridi}}, \bibinfo {author}
  {\bibfnamefont {E.}~\bibnamefont {Carretti}}, \bibinfo {author}
  {\bibfnamefont {E.}~\bibnamefont {Fratini}}, \bibinfo {author} {\bibfnamefont
  {A.}~\bibnamefont {Wiedenmann}}, \ and\ \bibinfo {author} {\bibfnamefont
  {P.}~\bibnamefont {Baglioni}},\ }\href@noop {} {\bibfield  {journal}
  {\bibinfo  {journal} {Langmuir}\ }\textbf {\bibinfo {volume} {24}},\ \bibinfo
  {pages} {12644} (\bibinfo {year} {2008})}\BibitemShut {NoStop}%
\bibitem [{\citenamefont {Fuhrer}\ \emph {et~al.}(2009)\citenamefont {Fuhrer},
  \citenamefont {Athanassiou}, \citenamefont {Luechinger},\ and\ \citenamefont
  {Stark}}]{fuhrer2009crosslinking}%
  \BibitemOpen
  \bibfield  {author} {\bibinfo {author} {\bibfnamefont {R.}~\bibnamefont
  {Fuhrer}}, \bibinfo {author} {\bibfnamefont {E.~K.}\ \bibnamefont
  {Athanassiou}}, \bibinfo {author} {\bibfnamefont {N.~A.}\ \bibnamefont
  {Luechinger}}, \ and\ \bibinfo {author} {\bibfnamefont {W.~J.}\ \bibnamefont
  {Stark}},\ }\href@noop {} {\bibfield  {journal} {\bibinfo  {journal} {Small}\
  }\textbf {\bibinfo {volume} {5}},\ \bibinfo {pages} {383} (\bibinfo {year}
  {2009})}\BibitemShut {NoStop}%
\bibitem [{\citenamefont {Frickel}\ \emph {et~al.}(2011)\citenamefont
  {Frickel}, \citenamefont {Messing},\ and\ \citenamefont
  {Schmidt}}]{frickel2011magneto}%
  \BibitemOpen
  \bibfield  {author} {\bibinfo {author} {\bibfnamefont {N.}~\bibnamefont
  {Frickel}}, \bibinfo {author} {\bibfnamefont {R.}~\bibnamefont {Messing}}, \
  and\ \bibinfo {author} {\bibfnamefont {A.~M.}\ \bibnamefont {Schmidt}},\
  }\href@noop {} {\bibfield  {journal} {\bibinfo  {journal} {J. Mater. Chem.}\
  }\textbf {\bibinfo {volume} {21}},\ \bibinfo {pages} {8466} (\bibinfo {year}
  {2011})}\BibitemShut {NoStop}%
\bibitem [{\citenamefont {Messing}\ \emph {et~al.}(2011)\citenamefont
  {Messing}, \citenamefont {Frickel}, \citenamefont {Belkoura}, \citenamefont
  {Strey}, \citenamefont {Rahn}, \citenamefont {Odenbach},\ and\ \citenamefont
  {Schmidt}}]{messing2011cobalt}%
  \BibitemOpen
  \bibfield  {author} {\bibinfo {author} {\bibfnamefont {R.}~\bibnamefont
  {Messing}}, \bibinfo {author} {\bibfnamefont {N.}~\bibnamefont {Frickel}},
  \bibinfo {author} {\bibfnamefont {L.}~\bibnamefont {Belkoura}}, \bibinfo
  {author} {\bibfnamefont {R.}~\bibnamefont {Strey}}, \bibinfo {author}
  {\bibfnamefont {H.}~\bibnamefont {Rahn}}, \bibinfo {author} {\bibfnamefont
  {S.}~\bibnamefont {Odenbach}}, \ and\ \bibinfo {author} {\bibfnamefont
  {A.~M.}\ \bibnamefont {Schmidt}},\ }\href@noop {} {\bibfield  {journal}
  {\bibinfo  {journal} {Macromolecules}\ }\textbf {\bibinfo {volume} {44}},\
  \bibinfo {pages} {2990} (\bibinfo {year} {2011})}\BibitemShut {NoStop}%
\bibitem [{\citenamefont {Weis}(2003)}]{weis2003simulation}%
  \BibitemOpen
  \bibfield  {author} {\bibinfo {author} {\bibfnamefont {J.~J.}\ \bibnamefont
  {Weis}},\ }\href@noop {} {\bibfield  {journal} {\bibinfo  {journal} {J.
  Phys.: Condens. Matter}\ }\textbf {\bibinfo {volume} {15}},\ \bibinfo {pages}
  {S1471} (\bibinfo {year} {2003})}\BibitemShut {NoStop}%
\bibitem [{\citenamefont {Weis}\ and\ \citenamefont
  {Levesque}(1993)}]{weis1993chain}%
  \BibitemOpen
  \bibfield  {author} {\bibinfo {author} {\bibfnamefont {J.~J.}\ \bibnamefont
  {Weis}}\ and\ \bibinfo {author} {\bibfnamefont {D.}~\bibnamefont
  {Levesque}},\ }\href@noop {} {\bibfield  {journal} {\bibinfo  {journal}
  {Phys. Rev. Lett.}\ }\textbf {\bibinfo {volume} {71}},\ \bibinfo {pages}
  {2729} (\bibinfo {year} {1993})}\BibitemShut {NoStop}%
\bibitem [{\citenamefont {Hynninen}\ and\ \citenamefont
  {Dijkstra}(2005)}]{hynninen2005phase}%
  \BibitemOpen
  \bibfield  {author} {\bibinfo {author} {\bibfnamefont {A.-P.}\ \bibnamefont
  {Hynninen}}\ and\ \bibinfo {author} {\bibfnamefont {M.}~\bibnamefont
  {Dijkstra}},\ }\href@noop {} {\bibfield  {journal} {\bibinfo  {journal}
  {Phys. Rev. E}\ }\textbf {\bibinfo {volume} {72}},\ \bibinfo {pages} {051402}
  (\bibinfo {year} {2005})}\BibitemShut {NoStop}%
\bibitem [{\citenamefont {Tavares}\ \emph {et~al.}(1999)\citenamefont
  {Tavares}, \citenamefont {Weis},\ and\ \citenamefont {Telo~da
  Gama}}]{tavares1999dipolar}%
  \BibitemOpen
  \bibfield  {author} {\bibinfo {author} {\bibfnamefont {J.~M.}\ \bibnamefont
  {Tavares}}, \bibinfo {author} {\bibfnamefont {J.~J.}\ \bibnamefont {Weis}}, \
  and\ \bibinfo {author} {\bibfnamefont {M.~M.}\ \bibnamefont {Telo~da Gama}},\
  }\href@noop {} {\bibfield  {journal} {\bibinfo  {journal} {Phys. Rev. E}\
  }\textbf {\bibinfo {volume} {59}},\ \bibinfo {pages} {4388} (\bibinfo {year}
  {1999})}\BibitemShut {NoStop}%
\bibitem [{\citenamefont {Klapp}\ and\ \citenamefont
  {Schoen}(2002)}]{klapp2002spontaneous}%
  \BibitemOpen
  \bibfield  {author} {\bibinfo {author} {\bibfnamefont {S.~H.~L.}\
  \bibnamefont {Klapp}}\ and\ \bibinfo {author} {\bibfnamefont
  {M.}~\bibnamefont {Schoen}},\ }\href@noop {} {\bibfield  {journal} {\bibinfo
  {journal} {J. Chem. Phys.}\ }\textbf {\bibinfo {volume} {117}},\ \bibinfo
  {pages} {8050} (\bibinfo {year} {2002})}\BibitemShut {NoStop}%
\bibitem [{\citenamefont {Klapp}(2005)}]{klapp2005dipolar}%
  \BibitemOpen
  \bibfield  {author} {\bibinfo {author} {\bibfnamefont {S.~H.~L.}\
  \bibnamefont {Klapp}},\ }\href@noop {} {\bibfield  {journal} {\bibinfo
  {journal} {J. Phys.: Condens. Matter}\ }\textbf {\bibinfo {volume} {17}},\
  \bibinfo {pages} {R525} (\bibinfo {year} {2005})}\BibitemShut {NoStop}%
\bibitem [{\citenamefont {Jarkova}\ \emph {et~al.}(2003)\citenamefont
  {Jarkova}, \citenamefont {Pleiner}, \citenamefont {M\"uller},\ and\
  \citenamefont {Brand}}]{jarkova2003hydrodynamics}%
  \BibitemOpen
  \bibfield  {author} {\bibinfo {author} {\bibfnamefont {E.}~\bibnamefont
  {Jarkova}}, \bibinfo {author} {\bibfnamefont {H.}~\bibnamefont {Pleiner}},
  \bibinfo {author} {\bibfnamefont {H.-W.}\ \bibnamefont {M\"uller}}, \ and\
  \bibinfo {author} {\bibfnamefont {H.~R.}\ \bibnamefont {Brand}},\ }\href@noop
  {} {\bibfield  {journal} {\bibinfo  {journal} {Phys. Rev. E}\ }\textbf
  {\bibinfo {volume} {68}},\ \bibinfo {pages} {041706} (\bibinfo {year}
  {2003})}\BibitemShut {NoStop}%
\bibitem [{\citenamefont {Bohlius}\ \emph {et~al.}(2004)\citenamefont
  {Bohlius}, \citenamefont {Brand},\ and\ \citenamefont
  {Pleiner}}]{bohlius2004macroscopic}%
  \BibitemOpen
  \bibfield  {author} {\bibinfo {author} {\bibfnamefont {S.}~\bibnamefont
  {Bohlius}}, \bibinfo {author} {\bibfnamefont {H.~R.}\ \bibnamefont {Brand}},
  \ and\ \bibinfo {author} {\bibfnamefont {H.}~\bibnamefont {Pleiner}},\
  }\href@noop {} {\bibfield  {journal} {\bibinfo  {journal} {Phys. Rev. E}\
  }\textbf {\bibinfo {volume} {70}},\ \bibinfo {pages} {061411} (\bibinfo
  {year} {2004})}\BibitemShut {NoStop}%
\bibitem [{\citenamefont {Bohlius}\ \emph {et~al.}(2007)\citenamefont
  {Bohlius}, \citenamefont {Pleiner},\ and\ \citenamefont
  {Brand}}]{bohlius2007solution}%
  \BibitemOpen
  \bibfield  {author} {\bibinfo {author} {\bibfnamefont {S.}~\bibnamefont
  {Bohlius}}, \bibinfo {author} {\bibfnamefont {H.}~\bibnamefont {Pleiner}}, \
  and\ \bibinfo {author} {\bibfnamefont {H.~R.}\ \bibnamefont {Brand}},\
  }\href@noop {} {\bibfield  {journal} {\bibinfo  {journal} {Phys. Fluids}\
  }\textbf {\bibinfo {volume} {19}} (\bibinfo {year} {2007})}\BibitemShut
  {NoStop}%
\bibitem [{\citenamefont {Stolbov}\ \emph {et~al.}(2011)\citenamefont
  {Stolbov}, \citenamefont {Raikher},\ and\ \citenamefont
  {Balasoiu}}]{stolbov2011modelling}%
  \BibitemOpen
  \bibfield  {author} {\bibinfo {author} {\bibfnamefont {O.~V.}\ \bibnamefont
  {Stolbov}}, \bibinfo {author} {\bibfnamefont {Y.~L.}\ \bibnamefont
  {Raikher}}, \ and\ \bibinfo {author} {\bibfnamefont {M.}~\bibnamefont
  {Balasoiu}},\ }\href@noop {} {\bibfield  {journal} {\bibinfo  {journal} {Soft
  Matter}\ }\textbf {\bibinfo {volume} {7}},\ \bibinfo {pages} {8484} (\bibinfo
  {year} {2011})}\BibitemShut {NoStop}%
\bibitem [{\citenamefont {Zubarev}(2012)}]{zubarev2012theory}%
  \BibitemOpen
  \bibfield  {author} {\bibinfo {author} {\bibfnamefont {A.~Y.}\ \bibnamefont
  {Zubarev}},\ }\href@noop {} {\bibfield  {journal} {\bibinfo  {journal} {Soft
  Matter}\ }\textbf {\bibinfo {volume} {8}},\ \bibinfo {pages} {3174} (\bibinfo
  {year} {2012})}\BibitemShut {NoStop}%
\bibitem [{\citenamefont {Wood}\ and\ \citenamefont
  {Camp}(2011)}]{camp2011modeling}%
  \BibitemOpen
  \bibfield  {author} {\bibinfo {author} {\bibfnamefont {D.~S.}\ \bibnamefont
  {Wood}}\ and\ \bibinfo {author} {\bibfnamefont {P.~J.}\ \bibnamefont
  {Camp}},\ }\href@noop {} {\bibfield  {journal} {\bibinfo  {journal} {Phys.
  Rev. E}\ }\textbf {\bibinfo {volume} {83}},\ \bibinfo {pages} {011402}
  (\bibinfo {year} {2011})}\BibitemShut {NoStop}%
\bibitem [{\citenamefont {Elfimova}\ \emph {et~al.}(2012)\citenamefont
  {Elfimova}, \citenamefont {Ivanov},\ and\ \citenamefont
  {Camp}}]{elfimova2012theory}%
  \BibitemOpen
  \bibfield  {author} {\bibinfo {author} {\bibfnamefont {E.~A.}\ \bibnamefont
  {Elfimova}}, \bibinfo {author} {\bibfnamefont {A.~O.}\ \bibnamefont
  {Ivanov}}, \ and\ \bibinfo {author} {\bibfnamefont {P.~J.}\ \bibnamefont
  {Camp}},\ }\href@noop {} {\bibfield  {journal} {\bibinfo  {journal} {J. Chem.
  Phys.}\ }\textbf {\bibinfo {volume} {136}},\ \bibinfo {pages} {194502}
  (\bibinfo {year} {2012})}\BibitemShut {NoStop}%
\bibitem [{\citenamefont {Ivaneyko}\ \emph {et~al.}(2011)\citenamefont
  {Ivaneyko}, \citenamefont {Toshchevikov}, \citenamefont {Saphiannikova},\
  and\ \citenamefont {Heinrich}}]{ivaneyko2011magneto}%
  \BibitemOpen
  \bibfield  {author} {\bibinfo {author} {\bibfnamefont {D.}~\bibnamefont
  {Ivaneyko}}, \bibinfo {author} {\bibfnamefont {V.~P.}\ \bibnamefont
  {Toshchevikov}}, \bibinfo {author} {\bibfnamefont {M.}~\bibnamefont
  {Saphiannikova}}, \ and\ \bibinfo {author} {\bibfnamefont {G.}~\bibnamefont
  {Heinrich}},\ }\href@noop {} {\bibfield  {journal} {\bibinfo  {journal}
  {Macromol. Theor. Simul.}\ }\textbf {\bibinfo {volume} {20}},\ \bibinfo
  {pages} {411} (\bibinfo {year} {2011})}\BibitemShut {NoStop}%
\bibitem [{\citenamefont {Dudek}\ \emph {et~al.}(2007)\citenamefont {Dudek},
  \citenamefont {Grabiec},\ and\ \citenamefont
  {Wojciechowski}}]{dudek2007molecular}%
  \BibitemOpen
  \bibfield  {author} {\bibinfo {author} {\bibfnamefont {M.~R.}\ \bibnamefont
  {Dudek}}, \bibinfo {author} {\bibfnamefont {B.}~\bibnamefont {Grabiec}}, \
  and\ \bibinfo {author} {\bibfnamefont {K.~W.}\ \bibnamefont
  {Wojciechowski}},\ }\href@noop {} {\bibfield  {journal} {\bibinfo  {journal}
  {Rev. Adv. Mater. Sci}\ }\textbf {\bibinfo {volume} {14}},\ \bibinfo {pages}
  {167} (\bibinfo {year} {2007})}\BibitemShut {NoStop}%
\bibitem [{\citenamefont {Weeber}\ \emph {et~al.}(2012)\citenamefont {Weeber},
  \citenamefont {Kantorovich},\ and\ \citenamefont
  {Holm}}]{weeber2012deformation}%
  \BibitemOpen
  \bibfield  {author} {\bibinfo {author} {\bibfnamefont {R.}~\bibnamefont
  {Weeber}}, \bibinfo {author} {\bibfnamefont {S.}~\bibnamefont {Kantorovich}},
  \ and\ \bibinfo {author} {\bibfnamefont {C.}~\bibnamefont {Holm}},\
  }\href@noop {} {\bibfield  {journal} {\bibinfo  {journal} {Soft Matter}\
  }\textbf {\bibinfo {volume} {8}},\ \bibinfo {pages} {9923} (\bibinfo {year}
  {2012})}\BibitemShut {NoStop}%
\bibitem [{\citenamefont {S\'anchez}\ \emph {et~al.}(2013)\citenamefont
  {S\'anchez}, \citenamefont {Cerd\`a}, \citenamefont {Sintes},\ and\
  \citenamefont {Holm}}]{sanchez2013filaments}%
  \BibitemOpen
  \bibfield  {author} {\bibinfo {author} {\bibfnamefont {P.~A.}\ \bibnamefont
  {S\'anchez}}, \bibinfo {author} {\bibfnamefont {J.~J.}\ \bibnamefont
  {Cerd\`a}}, \bibinfo {author} {\bibfnamefont {T.}~\bibnamefont {Sintes}}, \
  and\ \bibinfo {author} {\bibfnamefont {C.}~\bibnamefont {Holm}},\ }\href@noop
  {} {\bibfield  {journal} {\bibinfo  {journal} {arXiv}\ }\textbf {\bibinfo
  {volume} {cond-mat.mes-hall}},\ \bibinfo {pages} {1302.5845v1} (\bibinfo
  {year} {2013})}\BibitemShut {NoStop}%
\bibitem [{\citenamefont {Cerd\`a}\ \emph {et~al.}(2013)\citenamefont
  {Cerd\`a}, \citenamefont {S\'anchez}, \citenamefont {Holm},\ and\
  \citenamefont {Sintes}}]{cerda2013filaments}%
  \BibitemOpen
  \bibfield  {author} {\bibinfo {author} {\bibfnamefont {J.~J.}\ \bibnamefont
  {Cerd\`a}}, \bibinfo {author} {\bibfnamefont {P.~A.}\ \bibnamefont
  {S\'anchez}}, \bibinfo {author} {\bibfnamefont {C.}~\bibnamefont {Holm}}, \
  and\ \bibinfo {author} {\bibfnamefont {T.}~\bibnamefont {Sintes}},\
  }\href@noop {} {\bibfield  {journal} {\bibinfo  {journal} {arXiv}\ }\textbf
  {\bibinfo {volume} {cond-mat.soft}},\ \bibinfo {pages} {1302.5897v1}
  (\bibinfo {year} {2013})}\BibitemShut {NoStop}%
\bibitem [{\citenamefont {Percus}(1976)}]{Percus}%
  \BibitemOpen
  \bibfield  {author} {\bibinfo {author} {\bibfnamefont {J.~K.}\ \bibnamefont
  {Percus}},\ }\href@noop {} {\bibfield  {journal} {\bibinfo  {journal} {J.
  Stat. Phys.}\ }\textbf {\bibinfo {volume} {15}},\ \bibinfo {pages} {505}
  (\bibinfo {year} {1976})}\BibitemShut {NoStop}%
\bibitem [{\citenamefont {Rosenfeld}\ \emph {et~al.}(1997)\citenamefont
  {Rosenfeld}, \citenamefont {Schmidt}, \citenamefont {L\"owen},\ and\
  \citenamefont {Tarazona}}]{rosenfeld1997fundamental}%
  \BibitemOpen
  \bibfield  {author} {\bibinfo {author} {\bibfnamefont {Y.}~\bibnamefont
  {Rosenfeld}}, \bibinfo {author} {\bibfnamefont {M.}~\bibnamefont {Schmidt}},
  \bibinfo {author} {\bibfnamefont {H.}~\bibnamefont {L\"owen}}, \ and\
  \bibinfo {author} {\bibfnamefont {P.}~\bibnamefont {Tarazona}},\ }\href@noop
  {} {\bibfield  {journal} {\bibinfo  {journal} {Phys. Rev. E}\ }\textbf
  {\bibinfo {volume} {55}},\ \bibinfo {pages} {4245} (\bibinfo {year}
  {1997})}\BibitemShut {NoStop}%
\bibitem [{\citenamefont {Froltsov}\ \emph {et~al.}(2003)\citenamefont
  {Froltsov}, \citenamefont {Blaak}, \citenamefont {Likos},\ and\ \citenamefont
  {L{\"o}wen}}]{froltsov2003crystal}%
  \BibitemOpen
  \bibfield  {author} {\bibinfo {author} {\bibfnamefont {V.~A.}\ \bibnamefont
  {Froltsov}}, \bibinfo {author} {\bibfnamefont {R.}~\bibnamefont {Blaak}},
  \bibinfo {author} {\bibfnamefont {C.~N.}\ \bibnamefont {Likos}}, \ and\
  \bibinfo {author} {\bibfnamefont {H.}~\bibnamefont {L{\"o}wen}},\ }\href@noop
  {} {\bibfield  {journal} {\bibinfo  {journal} {Phys. Rev. E}\ }\textbf
  {\bibinfo {volume} {68}},\ \bibinfo {pages} {061406} (\bibinfo {year}
  {2003})}\BibitemShut {NoStop}%
\bibitem [{\citenamefont {Froltsov}\ \emph {et~al.}(2005)\citenamefont
  {Froltsov}, \citenamefont {Likos}, \citenamefont {L{\"o}wen}, \citenamefont
  {Eisenmann}, \citenamefont {Gasser}, \citenamefont {Keim},\ and\
  \citenamefont {Maret}}]{froltsov2005anisotropic}%
  \BibitemOpen
  \bibfield  {author} {\bibinfo {author} {\bibfnamefont {V.~A.}\ \bibnamefont
  {Froltsov}}, \bibinfo {author} {\bibfnamefont {C.~N.}\ \bibnamefont {Likos}},
  \bibinfo {author} {\bibfnamefont {H.}~\bibnamefont {L{\"o}wen}}, \bibinfo
  {author} {\bibfnamefont {C.}~\bibnamefont {Eisenmann}}, \bibinfo {author}
  {\bibfnamefont {U.}~\bibnamefont {Gasser}}, \bibinfo {author} {\bibfnamefont
  {P.}~\bibnamefont {Keim}}, \ and\ \bibinfo {author} {\bibfnamefont
  {G.}~\bibnamefont {Maret}},\ }\href@noop {} {\bibfield  {journal} {\bibinfo
  {journal} {Phys. Rev. E}\ }\textbf {\bibinfo {volume} {71}},\ \bibinfo
  {pages} {031404} (\bibinfo {year} {2005})}\BibitemShut {NoStop}%
\bibitem [{\citenamefont {de~Gennes}(1980)}]{degennes1980}%
  \BibitemOpen
  \bibfield  {author} {\bibinfo {author} {\bibfnamefont {P.~G.}\ \bibnamefont
  {de~Gennes}},\ }in\ \href@noop {} { {\bibinfo {booktitle} \textit{Liquid
  crystals of one- and two-dimensional order}}},\ \bibinfo {editor} {edited by\
  \bibinfo {editor} {\bibfnamefont {W.}~\bibnamefont {Helfrich}}\ and\ \bibinfo
  {editor} {\bibfnamefont {G.}~\bibnamefont {Heppke}}}\ (\bibinfo  {publisher}
  {Springer, Berlin},\ \bibinfo {year} {1980})\ p.\ \bibinfo {pages} {231
  ff.}\BibitemShut {Stop}%
\bibitem [{\citenamefont {Brand}\ and\ \citenamefont
  {Pleiner}(1994)}]{brand1994electrohydrodynamics}%
  \BibitemOpen
  \bibfield  {author} {\bibinfo {author} {\bibfnamefont {H.~R.}\ \bibnamefont
  {Brand}}\ and\ \bibinfo {author} {\bibfnamefont {H.}~\bibnamefont
  {Pleiner}},\ }\href@noop {} {\bibfield  {journal} {\bibinfo  {journal}
  {Physica A}\ }\textbf {\bibinfo {volume} {208}},\ \bibinfo {pages} {359}
  (\bibinfo {year} {1994})}\BibitemShut {NoStop}%
\bibitem [{\citenamefont {Menzel}\ \emph {et~al.}(2007)\citenamefont {Menzel},
  \citenamefont {Pleiner},\ and\ \citenamefont {Brand}}]{menzel2007nonlinear}%
  \BibitemOpen
  \bibfield  {author} {\bibinfo {author} {\bibfnamefont {A.~M.}\ \bibnamefont
  {Menzel}}, \bibinfo {author} {\bibfnamefont {H.}~\bibnamefont {Pleiner}}, \
  and\ \bibinfo {author} {\bibfnamefont {H.~R.}\ \bibnamefont {Brand}},\
  }\href@noop {} {\bibfield  {journal} {\bibinfo  {journal} {J. Chem. Phys.}\
  }\textbf {\bibinfo {volume} {126}},\ \bibinfo {pages} {234901} (\bibinfo
  {year} {2007})}\BibitemShut {NoStop}%
\bibitem [{\citenamefont {Menzel}\ \emph
  {et~al.}(2009{\natexlab{a}})\citenamefont {Menzel}, \citenamefont {Pleiner},\
  and\ \citenamefont {Brand}}]{menzel2009nonlinear}%
  \BibitemOpen
  \bibfield  {author} {\bibinfo {author} {\bibfnamefont {A.~M.}\ \bibnamefont
  {Menzel}}, \bibinfo {author} {\bibfnamefont {H.}~\bibnamefont {Pleiner}}, \
  and\ \bibinfo {author} {\bibfnamefont {H.~R.}\ \bibnamefont {Brand}},\
  }\href@noop {} {\bibfield  {journal} {\bibinfo  {journal} {J. Appl. Phys.}\
  }\textbf {\bibinfo {volume} {105}},\ \bibinfo {pages} {013503} (\bibinfo
  {year} {2009}{\natexlab{a}})}\BibitemShut {NoStop}%
\bibitem [{\citenamefont {Menzel}\ \emph
  {et~al.}(2009{\natexlab{b}})\citenamefont {Menzel}, \citenamefont {Pleiner},\
  and\ \citenamefont {Brand}}]{menzel2009response}%
  \BibitemOpen
  \bibfield  {author} {\bibinfo {author} {\bibfnamefont {A.~M.}\ \bibnamefont
  {Menzel}}, \bibinfo {author} {\bibfnamefont {H.}~\bibnamefont {Pleiner}}, \
  and\ \bibinfo {author} {\bibfnamefont {H.~R.}\ \bibnamefont {Brand}},\
  }\href@noop {} {\bibfield  {journal} {\bibinfo  {journal} {Eur. Phys. J. E}\
  }\textbf {\bibinfo {volume} {30}},\ \bibinfo {pages} {371} (\bibinfo {year}
  {2009}{\natexlab{b}})}\BibitemShut {NoStop}%
\bibitem [{\citenamefont {Brand}\ \emph {et~al.}(2011)\citenamefont {Brand},
  \citenamefont {Martinoty},\ and\ \citenamefont
  {Pleiner}}]{brand2011physical}%
  \BibitemOpen
  \bibfield  {author} {\bibinfo {author} {\bibfnamefont {H.~R.}\ \bibnamefont
  {Brand}}, \bibinfo {author} {\bibfnamefont {P.}~\bibnamefont {Martinoty}}, \
  and\ \bibinfo {author} {\bibfnamefont {H.}~\bibnamefont {Pleiner}},\ }in\
  \href@noop {} { {\bibinfo {booktitle} \textit{Cross-linked liquid crystalline
  systems: from rigid polymer networks to elastomers}}},\ \bibinfo {series and
  number} {The liquid crystals book series},\ \bibinfo {editor} {edited by\
  \bibinfo {editor} {\bibfnamefont {D.}~\bibnamefont {Broer}}, \bibinfo
  {editor} {\bibfnamefont {G.}~\bibnamefont {Crawford}}, \ and\ \bibinfo
  {editor} {\bibfnamefont {S.}~\bibnamefont {Zumer}}}\ (\bibinfo  {publisher}
  {CRC Press Inc},\ \bibinfo {year} {2011})\ pp.\ \bibinfo {pages}
  {529--563}\BibitemShut {NoStop}%
\bibitem [{\citenamefont {Miller}\ \emph {et~al.}(2009)\citenamefont {Miller},
  \citenamefont {Blaak}, \citenamefont {Lumb},\ and\ \citenamefont
  {Hansen}}]{miller2009dynamical}%
  \BibitemOpen
  \bibfield  {author} {\bibinfo {author} {\bibfnamefont {M.~A.}\ \bibnamefont
  {Miller}}, \bibinfo {author} {\bibfnamefont {R.}~\bibnamefont {Blaak}},
  \bibinfo {author} {\bibfnamefont {C.~N.}\ \bibnamefont {Lumb}}, \ and\
  \bibinfo {author} {\bibfnamefont {J.-P.}\ \bibnamefont {Hansen}},\
  }\href@noop {} {\bibfield  {journal} {\bibinfo  {journal} {J. Chem. Phys.}\
  }\textbf {\bibinfo {volume} {130}},\ \bibinfo {pages} {114507} (\bibinfo
  {year} {2009})}\BibitemShut {NoStop}%
\bibitem [{\citenamefont {Dobrushin}(1973)}]{dobrushin1973analyticity}%
  \BibitemOpen
  \bibfield  {author} {\bibinfo {author} {\bibfnamefont {R.~L.}\ \bibnamefont
  {Dobrushin}},\ }\href@noop {} {\bibfield  {journal} {\bibinfo  {journal}
  {Commun. Math. Phys.}\ }\textbf {\bibinfo {volume} {32}},\ \bibinfo {pages}
  {269} (\bibinfo {year} {1973})}\BibitemShut {NoStop}%
\bibitem [{\citenamefont {Fr{\"o}hlich}\ and\ \citenamefont
  {Spencer}(1982)}]{frohlich1982phase}%
  \BibitemOpen
  \bibfield  {author} {\bibinfo {author} {\bibfnamefont {J.}~\bibnamefont
  {Fr{\"o}hlich}}\ and\ \bibinfo {author} {\bibfnamefont {T.}~\bibnamefont
  {Spencer}},\ }\href@noop {} {\bibfield  {journal} {\bibinfo  {journal}
  {Commun. Math. Phys.}\ }\textbf {\bibinfo {volume} {84}},\ \bibinfo {pages}
  {87} (\bibinfo {year} {1982})}\BibitemShut {NoStop}%
\bibitem [{\citenamefont {Strobl}(2007)}]{strobl1997physics}%
  \BibitemOpen
  \bibfield  {author} {\bibinfo {author} {\bibfnamefont {G.}~\bibnamefont
  {Strobl}},\ }\href@noop {} { {\bibinfo {title} \textit{The physics of
  polymers}}}\ (\bibinfo  {publisher} {Springer, Berlin Heidelberg},\ \bibinfo
  {year} {2007})\BibitemShut {NoStop}%
\bibitem [{\citenamefont {Ivaneyko}\ \emph {et~al.}(2012)\citenamefont
  {Ivaneyko}, \citenamefont {Toshchevikov}, \citenamefont {Saphiannikova},\
  and\ \citenamefont {Heinrich}}]{ivaneyko2012effects}%
  \BibitemOpen
  \bibfield  {author} {\bibinfo {author} {\bibfnamefont {D.}~\bibnamefont
  {Ivaneyko}}, \bibinfo {author} {\bibfnamefont {V.}~\bibnamefont
  {Toshchevikov}}, \bibinfo {author} {\bibfnamefont {M.}~\bibnamefont
  {Saphiannikova}}, \ and\ \bibinfo {author} {\bibfnamefont {G.}~\bibnamefont
  {Heinrich}},\ }\href@noop {} {\bibfield  {journal} {\bibinfo  {journal}
  {Condens. Matter Phys.}\ }\textbf {\bibinfo {volume} {15}},\ \bibinfo {pages}
  {33601} (\bibinfo {year} {2012})}\BibitemShut {NoStop}%
\bibitem [{\citenamefont {Han}\ \emph {et~al.}(2013)\citenamefont {Han},
  \citenamefont {Hong},\ and\ \citenamefont {Faidley}}]{han2013field}%
  \BibitemOpen
  \bibfield  {author} {\bibinfo {author} {\bibfnamefont {Y.}~\bibnamefont
  {Han}}, \bibinfo {author} {\bibfnamefont {W.}~\bibnamefont {Hong}}, \ and\
  \bibinfo {author} {\bibfnamefont {L.~E.}\ \bibnamefont {Faidley}},\
  }\href@noop {} {\bibfield  {journal} {\bibinfo  {journal} {Int. J. Solids
  Struct.}\ }\textbf {\bibinfo {volume} {50}},\ \bibinfo {pages}
  {2281}  (\bibinfo {year} {2013})}\BibitemShut {NoStop}%
\end{thebibliography}
\end{document}